\documentclass[fleqn, usenatbib]{mnras}

\DeclareRobustCommand{\VAN}[3]{#2}
\let\VANthebibliography\thebibliography
\def\thebibliography{\DeclareRobustCommand{\VAN}[3]{##3}\VANthebibliography}

\usepackage{graphicx}	
\usepackage{amsmath}	
\usepackage{amssymb}	
\usepackage{bm}			
\usepackage{color}

\newcommand{\vs}{\vspace}

\renewcommand{\.}{\hspace{0.5mm}}
\newcommand{\Mrm}{\ensuremath{\mathrm{M}}}
\newcommand{\Hrm}{\ensuremath{\mathrm{H}}}
\newcommand{\Srm}{\ensuremath{\mathrm{S}}}

\newcommand{\Ocal}{\ensuremath{\mathcal{O}}}
\newcommand{\Crm}{\ensuremath{\mathrm{C}}}
\newcommand{\Erm}{\ensuremath{\mathrm{E}}}
\newcommand{\Brm}{\ensuremath{\mathrm{B}}}
\newcommand{\mrm}{\ensuremath{\mathrm{m}}}

\newcommand{\rrm}{\ensuremath{\mathrm{r}}}
\newcommand{\Krm}{\ensuremath{\mathrm{K}}}

\newcommand{\prm}{\ensuremath{\mathrm{p}}}

\newcommand{\Frm}{\ensuremath{\mathrm{F}}}

\newcommand{\sv}{\langle\sigma v \rangle}

\newcommand{\Lrm}{\ensuremath{\mathrm{L}}}
\def\fPBH{f_{\rm PBH}}

\definecolor{orange}{rgb}{1,0.5,0}

\renewcommand\({\left(}
\renewcommand\){\right)}
\renewcommand\[{\left[}
\renewcommand\]{\right]}
\renewcommand{\d}{\ensuremath{\mathrm{d}}}

\title{Black Holes and WIMPs: All or Nothing or Something Else}

\author[B. Carr et al.]{
Bernard Carr,$^{1, \,2}$\thanks{E-mail: B.J.Carr@qmul.ac.uk}
Florian K{\"u}hnel,$^{3}$\thanks{E-mail: kuhnel@kth.se}
and Luca Visinelli$^{4, \,5}$\thanks{E-mail: luca.visinelli@sjtu.edu.cn; ``Fellini'' Marie Curie fellow}
\\
$^{1}$School of Physics and Astronomy, 
		Queen Mary University of London,
		Mile End Road, 
		London E1 4NS, 
		UK\\
$^{2}$Research Center for the Early Universe, 
		University of Tokyo, 
		Tokyo 113-0033, 
		Japan\\
$^{3}$Arnold Sommerfeld Center,
		Ludwig-Maximilians-Universit{\"a}t,
		Theresienstra{\ss}e 37,
		80333 M{\"u}nchen,
		Germany\\
$^{4}$Gravitation Astroparticle Physics Amsterdam (GRAPPA),\\
		Institute for Theoretical Physics Amsterdam and Delta Institute for Theoretical Physics,\\
		University of Amsterdam, Science Park 904, 1098 XH Amsterdam, The Netherlands\\
$^{5}$INFN, Laboratori Nazionali di Frascati, C.P. 13, 100044 Frascati, Italy
}

\date{Accepted XXX. Received YYY; in original form ZZZ}
\pubyear{2020}

\begin{document}
\label{firstpage}
\pagerange{\pageref{firstpage}--\pageref{lastpage}}
\maketitle

\begin{abstract}
We consider constraints on primordial black holes (PBHs) in the mass range $( 10^{-18}\text{--}10^{15} )\,M_{\odot}$ if the dark matter (DM) comprises weakly interacting massive particles (WIMPs) which form halos around them and generate $\gamma$-rays by annihilations. We first study the formation of the halos and find that their density profile prior to WIMP annihilations evolves to a characteristic power-law form. Because of the wide range of PBH masses considered, our analysis forges an interesting link between previous approaches to this problem. We then consider the effect of the WIMP annihilations on the halo profile and the associated generation of $\gamma$-rays. The observed extragalactic $\gamma$-ray background implies that the PBH DM fraction is $f^{}_{\rm PBH} \lesssim 2 \times 10^{-9}\,( m_{\chi} / {\rm TeV} )^{1.1}$ in the mass range $2 \times 10^{-12}\,M_{\odot}\,( m_{\chi} / {\rm TeV} )^{-3.2} \lesssim M \lesssim 5 \times 10^{12}\,M_{\odot}\,( m_{\chi} / {\rm TeV} )^{1.1}$, where $m_{\chi}$ and $M$ are the WIMP and PBH masses, respectively. This limit is independent of $M$ and therefore applies for any PBH mass function. For $M \lesssim 2\times 10^{-12}\,M_{\odot}\,( m_{\chi}/ {\rm TeV} )^{-3.2}$, the constraint on $f^{}_{\rm PBH}$ is a decreasing function of $M$ and PBHs could still make a significant DM contribution at very low masses. We also consider constraints on WIMPs if the DM is mostly PBHs. If the merging black holes recently discovered by LIGO/Virgo are of primordial origin, this would rule out the standard WIMP DM scenario. More generally, the WIMP DM fraction cannot exceed $10^{-4}$ for $M > 10^{-9}\,M_{\odot}$ and $m_{\chi} > 10\,$GeV. There is a region of parameter space, with $M \lesssim 10^{-11}\,M_{\odot}$ and $m_{\chi} \lesssim 100\,$GeV, in which WIMPs and PBHs can both provide some but not all of the DM, so that one requires a third DM candidate.
\end{abstract}


\section{Introduction}
\label{sec:Introduction}

The recent discovery of intermediate-mass black-hole mergers by the LIGO/Virgo collaboration~\citep{PhysRevLett.125.101102} has led to speculation that the dark matter (DM) might consist of black holes rather than a more conventional candidate, such as a weakly interacting massive particle (WIMP). This is due to the of the LIGO/Virgo black holes being produced through stellar collapse or multi-stage mergers. Although the LIGO/Virgo black holes might not be numerous enough to explain {\it all} the DM, they would need to provide at least $1\%$ of it, which suggests the possibility of a hybrid model, in which the DM is some mixture of WIMPs and black holes.

If black holes provide more than $20\%$ of the dark matter, the success of the cosmological nucleosynthesis scenario~\citep{1967ApJ...148....3W} implies they could not derive from baryons and would therefore need to be {\it primordial} in origin. The suggestion that the DM could be primordial black holes (PBHs) dates back to the 1970s~\citep{Carr:1974nx, 1975Natur.253..251C} but has intensified over the past three decades, partly due to the failure to find either experimental or astronomical evidence for WIMPs. If PBHs have monochromatic mass function, there are only a few mass windows in which they could provide all the DM but the situation is more complicated in the realistic case in which they have an extended mass function~\citep{Kuhnel:2015vtw, Carr:2016drx, Kuhnel:2017pwq, Carr:2017jsz}. For example, one would expect the mass at which the density peaks to be less than the mass at which the LIGO/Virgo events peak, since the gravitational wave signal is stronger for more massive PBHs. In particular, it has been pointed out that the thermal history of the Universe may naturally generate a bumpy PBH mass function, which could explain the DM, the LIGO/Virgo events and various other cosmological conundra~\citep{Carr:2019kxo}. For a recent comprehensive review of these issues, see~\cite{Carr:2020xqk}.

Whether or not the black holes are primordial{\;---\;}and the analysis of this paper will cover both cases{\;---\;}there is a serious objection to hybrid models in which most of the DM comprises WIMPs. This is because they would inevitably clump in halos around the black holes, generating enhanced annihilations and $\gamma$-ray emission. As first studied by~\cite{Mack:2006gz, Ricotti:2007jk, Ricotti:2007au, Ricotti:2009bs, Lacki:2010zf} and, more recently, by~\cite{Eroshenko:2016yve, Boucenna:2017ghj, Adamek:2019gns, Bertone:2019vsk, Eroshenko:2019pxt, Cai:2020fnq}, this implies very stringent constraints on hybrid scenarios, leading to the conclusion that one cannot have an appreciable amount of DM in {\it both} components.\footnote{The title of our paper is inspired by~\cite{Lacki:2010zf}, who also considered constraints on the WIMP parameters.} If nearly all the DM is WIMPs, the fraction in black holes must be tiny; but if nearly all the DM is PBHs, the fraction in WIMPs must be tiny.

However, this problem has only been investigated for a rather restricted combination of black hole and WIMP masses, so the previous analysis needs to be extended to see if this conclusion applies more generally. For example,~\cite{Adamek:2019gns} focus on the PBH mass range around $1\,M_{\odot}$ in which the WIMP velocity distribution can be neglected but allow a range of WIMP masses ($10\,$GeV to $1\,$TeV);~\cite{Eroshenko:2016yve} focuses on the subsolar PBH mass range where the velocity distribution must be included but assume a particular WIMP mass ($70\,$GeV). \cite{Kadota:2020ahr} discuss the annihilation of sub-GeV dark matter around PBHs. The present analysis uses a combination of numerical and analytical techniques to amalgamate these three approaches and elucidates the connection between them.

We must also distinguish between the Galactic and extragalactic $\gamma$-ray backgrounds associated with WIMP annihilations, the constraint associated with latter being stronger for all PBH and WIMP masses. There has also been a study of the annihilation signal from the halo of WIMPs around the supermassive black hole in the Galactic centre~\citep{Hooper:2010mq} and we stress that Ultracompact Minihalos may be associated with WIMP annihilations, even if they do not contain a central black hole~\citep{Scott:2009tu, Josan:2010vn, Bringmann:2011ut}. However, we do not consider these cases here.

In a recent paper~\citep{Carr:2020erq}, we have studied {\it stupendously large} black holes (SLABs) in the mass range $10^{11}\text{--}10^{18}\,M_{\odot}$. Such enormous objects might conceivably reside in galactic nuclei, since there is already evidence for black holes of up to $7 \times 10^{10}\,M_{\odot}$~\citep{Shemmer:2004ph} in that context. However, our considerations were mainly motivated by the apparent lack of constraints on PBHs in this mass range. Although SLABs are obviously too large to provide the DM in galactic halos, they might still have a large cosmological density. We found that the accretion constraints in this mass range are beset with astrophysical uncertainties, so the WIMP annihilation limit is the cleanest, at least if WIMPs provide most of the DM. The strongest limit then comes from the extragalactic $\gamma$-ray background and the constraint on the DM fraction is independent of the black hole mass.

At the other extreme, it is interesting to consider the possibility of stupendously {\it small} black holes, since there is still a window in the sub-planetary (asteroid to lunar) mass range ($10^{-16}\text{--}10^{-10}\,M_{\odot}$ or $10^{17}\text{--}10^{23}\,$g) where PBHs could provide the DM. Since these are much smaller $1\,M_{\odot}$, they are necessarily primordial, so there is no longer the ambiguity associated with SLABs. The WIMP-annihilation constraints become weaker for lighter black holes, so we need to determine whether a scenario in which both WIMPs and sub-planetary PBHs have an appreciable density is necessarily excluded. If the sum of their contributions were less than $100\%$, one would be forced to a scenario which involves a {\it third} DM candidate.

To fill the gap in the previous literature, the purpose of this paper is to study the interplay between the DM candidates over the PBH mass range $10^{-18}$ to $10^{15}\,M_{\odot}$ and the WIMP mass range $10\,{\rm GeV}\text{--}1\,{\rm TeV}$. Section~\ref{sec:Thermal-Production-of-WIMPs} discusses the thermal production of WIMPs. Section \ref{sec:Structure-of-the-Dark--Matter-Halos} determines the structure of the resulting DM halos. Section~\ref{sec:Gamma--Ray-Flux-from-WIMP-Annihilation} derives the the Galactic and extragalactic $\gamma$-ray flux from WIMP annihilations in these halos. Section \ref{sec:Implications-from-Detections} discusses the implications of the recent LIGO/Virgo gravitational-wave events are due to merging PBHs. Section~\ref{sec:Discusson-and-Outlook} concludes with a discussion of future prospects. Throughout this paper we choose units with $c = k_{\Brm} = 1$. The code used to produce the results for this work is publicly available at \href{https://github.com/lucavisinelli/WIMPdistributionPBH}{github.com/lucavisinelli/WIMPdistributionPBH.}

\section{Thermal Production of WIMPs}
\label{sec:Thermal-Production-of-WIMPs}

In the following, we assume that WIMPs are their own antiparticles and that Maxwell-Boltzmann statistics suffices in describing the distribution of the particles. At temperatures much higher than the WIMP mass $m_{\chi}$, the production of WIMPs in the primordial plasma of the early Universe proceeds through particle-antiparticle collisions, with rate
\begin{equation}
	\label{eq:gamma-ann}
	\Gamma_{\rm ann}
		=
					\langle \sigma v \rangle^{}_{\rm th}\,n_{\rm eq}
					\, .
\end{equation}
Here, $\sigma$ is the WIMP annihilation cross section, $v$ is the WIMP relative velocity, angle brackets denote an average over the WIMP thermal distribution (th), and $n$ is the WIMP number density, with the value $n_{\rm eq}$ for chemical equilibrium. In more detail, the yield $Y \equiv n / s$ in terms of the entropy density $s$, evolves as~\citep{Lee:1977ua, Steigman:1979kw}
\begin{equation}
	\frac{\d Y}{\d x}
		=
					\frac{ 1 }{ 3 H }\.\frac{ \d s }{ \d x }\.
					\langle \sigma v \rangle^{}_{\rm th}
					\(
						Y^{2}
						-
						Y_{\rm eq}^{2}
					\)
					,
\end{equation}
where $Y_{\rm eq} \equiv n_{\rm eq} / s$ and the independent variable $x \equiv m_{\chi} / T$ is inversely proportional to the plasma temperature $T$. This equation assumes the conservation of entropy in a comoving volume throughout its range of applicability. It can be solved numerically with the initial condition $Y = Y_{\rm eq}$ at $x \approx 1$ to obtain the present yield $Y_{0}$ and the WIMP relic density,
\begin{equation}
	\label{eq:WIMP-fraction}
	\Omega_{\chi}\,h^{2}
		=
					\frac{ m_{\chi}\,s_{0}\,Y_{0}\,h^{2} }
					{ \rho_{\rm crit} }
					\, ,
\end{equation}
where $s_{0}$ is the entropy density at present time, $h \equiv H_{0} / \( 100{\rm \,km\,s^{-1}\,Mpc^{-1}} \)$ and $\rho_{\rm crit} = 3 H_{0}^{2} / ( 8\pi\.G )$ is the present value of the critical density.

The annihilation rate governs the Boltzmann equation that tracks the WIMP number density. At a temperature $T_{\Frm} \approx m_{\chi} / 20$, the annihilation rate falls below the cosmic expansion rate, so the WIMPs cease to be produced and {\it chemically} decouple. For $T \lesssim T_{\Frm}$, the number of WIMPs in a comoving volume remains approximately constant until the present.

The computation of $Y_{0}$ strongly depends on the value of $\langle \sigma v \rangle^{}_{\rm th}$. If we can neglect co-annihilation~\citep{Griest:1990kh} and Sommerfeld enhancement~\citep{ArkaniHamed:2008qn}, we obtain~\citep{Gondolo:1990dk}
\begin{equation}
	\label{eq:sigmavGondolo}
	\langle \sigma v \rangle^{}_{\rm th}
		=
					\frac{\int_{4 m_{\chi}^{2}}^{+\infty}\d s
					\(
						s - 4\.m_{\chi}^{2}
					\)
					\sqrt{s\,}\,K_{1}( \sqrt{s\,} / T )\,
					\sigma( s )}
					{8\,m_{\chi}^{4}\,T
					\[
						K_{2}( m_{\chi} / T )
					\]^{2}}
					\, ,
\end{equation}
where $s$ is the center-of-mass energy squared and $K_{\nu}$ is the modified Bessel function of the second kind of order $\nu$. If the product $\sigma v$ varies slowly with $v$, it can be approximated as
\begin{equation}
	\label{eq:taylor}
	\sigma v
		\simeq
					a + b\.v^{2}
					\, ,
\end{equation}
with constants $a$ and $b$, so that $\langle \sigma v \rangle^{}_{\rm th} = a + 3\.b\.T / ( 2 m_{\chi} )$. At lowest order in the non-relativistic expansion, $\langle \sigma_{\rm ann }v \rangle^{}_{\rm th}$ is then independent of the WIMP velocity distribution and the same throughout the history of the Universe.

Near resonances and thresholds where $\sigma v$ varies rapidly with $v$, such an expansion is no longer valid, and Eq.~\eqref{eq:sigmavGondolo} has to be implemented more carefully. Co-annihilation can be handled in realistic models with numerical packages~\citep{Bringmann:2018lay, Belanger:2018ccd}. In the following, we assume that chemical decoupling occurs in the radiation-dominated period of the standard model. However, one can also consider production in non-standard cosmological scenarios~\citep{Gelmini:2006pw, Acharya:2009zt, Visinelli:2017qga}.

Following the methods outlined by~\cite{Steigman:2012nb} and~\cite{Baum:2016oow}, we compute the expression for $\langle \sigma v \rangle^{}_{\rm th}$ that produces a WIMP fraction $f_{\chi} = \Omega_{\chi} / \Omega_{\rm DM}$ through thermal freeze-out~\citep{Lee:1977ua, Hut:1977zn, Sato:1977ye}. To lowest order in $T / m_{\chi}$, a numerical fit gives
\begin{equation}
	\label{eq:sigmavth}
	\langle \sigma v \rangle^{}_{\rm th}
		=
					\langle \sigma v \rangle^{}_{\rm DM}\,
					f_{\chi}^{-1.0}
					\, ,
\end{equation}
where $\langle \sigma v \rangle^{}_{\rm DM} = 2.5 \times 10^{-26}{\rm\,cm^{3}\,s^{-1}}$. This expression is valid over a wide range of WIMP densities and masses above $10\,$GeV. The only deviation from the $\langle \sigma v \rangle^{}_{\rm th} \propto f_{\chi}^{-1}$ behaviour is due to the changes in the relativistic degrees of freedom with temperature.

At temperatures below $T_{\Frm}$, the relativistic plasma and WIMPs continue to exchange energy and momentum even if the comoving number of WIMPs is fixed at the value $n_{\rm eq}$. This applies as long as the Hubble rate is higher than the WIMP scattering rate~\citep{Bernstein:1985th}, after which the WIMPs also decouple {\it kinetically}. Kinetic decoupling (KD) occurs at the temperature~\citep{Bringmann:2006mu, Visinelli:2015eka}
\begin{equation}
	T_{\rm KD}
		=
					\frac{ m_{\chi}}{ \Gamma(3/4) }\mspace{-2mu}
					\(
						\frac{ g\,m_{\chi} }{ M_{\rm Pl} }
					\)^{\!1/4}
		\sim
					10 {\rm\, MeV}
					\(
						\frac{m_{\chi} }{100\,{\rm GeV} }
					\)^{5/4}
					\, ,
					\label{eq:KDtemp}
\end{equation}
where $g \approx 10.9$ for temperatures in the range $0.1\text{--}10\,$MeV and $\Gamma( 3/4 ) \approx 1.225$. The corresponding Hubble rate and time are $H_{\rm KD}$ and $t_{\rm KD} = 1 / ( 2\.H_{\rm KD} )$, respectively. Since the KD temperature is much smaller than the WIMP mass, and since WIMPs are in thermal equilibrium with the plasma down to $T_{\rm KD}$ due to scatterings off SM particles, we assume they
have a Boltzmann velocity distribution.

\section{Structure of Dark-Matter Halos}
\label{sec:Structure-of-the-Dark--Matter-Halos}

PBHs form during the radiation-dominated epoch from the collapse of mildly non-linear perturbations. 
It is well known that the presence of a black hole can lead to a {\it spike} in the distribution of the surrounding WIMPs. Indeed, non-relativistic WIMPs produced after freeze-out can already be gravitationally bound to PBHs at their formation~\citep{Ricotti:2007au, Lacki:2010zf, Saito:2010ts, Xu:2020jpv}. Thereafter more WIMPs will be gravitationally attracted to them, leading to the formation of halos whose profile depends on the WIMP velocity distribution. However, as discussed below, the spike distribution will be flattened sufficiently close to the black hole by either DM scattering off plasma prior to KD or DM annihilation.

\subsection{Halo formation before matter-radiation equality}

A WIMP at a distance $r$ from a PBH of mass $M$ experiences a gravitational attraction $\sim G M / r^{2}$ and a cosmic deceleration $\sim H^{2}r$. The turn-around of the WIMP orbit occurs where these two terms are comparable at a radius $\sim ( G M / H^{2} )^{1/3}$ with $H = 1 / ( 2\.t )$ during radiation-domination. Indeed, a detailed numerical solution for the WIMP equation of motion shows that the turn-around radius is well approximated by~\citep{Adamek:2019gns}
\begin{equation}
	\label{eq:turnaround}
	r_{\rm ta}( t )
		\approx 1.0
					\left(
						r_{\Srm}\.t^{2}
					\right)^{1/3}
					\, ,
\end{equation}
where $r_{\Srm} = 2\.G M$. Since a PBH has around the cosmological horizon size $\sim t$ at formation, the radius $r_{\rm ta}$ necessarily exceeds $r_{\Srm}$ thereafter and it is always smaller than the cosmological horizon. The mass within $r_{\rm ta}$ remains comparable to the PBH mass until the time of matter-radiation equality, $t_{\rm eq}$. Indeed, $r_{\rm ta}$ can be regarded as the radius of influence of the black hole, in the sense that it contains the same mass as the black hole~\citep{Eroshenko:2016yve}. However, the mass in WIMPs is only some fraction of this before $t_{\rm eq}$.

Let us first consider the effect of PBHs which form before $t_{\rm KD}$. From Eq.~\eqref{eq:KDtemp}, these must be lighter than
\begin{equation}
	M_{\rm KD}
		\sim
					10^{3}
					\(
						\frac{ m_{\chi} }{ 100\,{\rm GeV} }
					\)^{-5/2}
					M_{\odot}
					\, ,
\end{equation}
which covers most of the mass range in which PBHs are usually invoked. Few WIMPs are captured by the PBH prior to KD since they are tightly coupled to the primordial plasma (mainly radiation) and there is very little accretion of radiation before $t_{\rm eq}$~\citep{Carr:1974nx}. However, a PBH that has already formed at KD would be surrounded by a halo of WIMPs up to the radius $r_{\rm ta}( t_{\rm KD})$ with a uniform density profile corresponding to the background WIMP density at that epoch, $\rho_{\rm KD}$.

After $t_{\rm KD}$ the halo radius will grow but the density will fall beyond $r_{\rm ta}( t_{\rm KD} )$. Since the halo density at radius $r$ is close to the background WIMP density ($\propto a^{-3}$) when this is the turn-around radius, the density profile is $\rho ( r ) \propto t^{-3/2} \propto r^{-9/4}$ using Eq.~\eqref{eq:turnaround}. The WIMP halo therefore steadily grows until $t_{\rm eq}$, at which point the profile has the spike form~\citep{Adamek:2019gns}
\begin{align}
\begin{split}
	\label{eq:densityprofile}
	\rho_{\rm \chi, \,spike}( r ) 
		 &= 
					f_{\chi}\.\frac{ \rho_{\rm eq} }{ 2 }\!
					\left(
						\frac{ r_{\rm ta}( t_{\rm eq} ) }{ r }
					\right)^{\!9/4}
					\quad
					[
						r_{\rm ta}(t_{\rm KD})
						<
						r
						<
						r_{\rm ta}( t_{\rm eq} )
					]
					\\[1mm]
		 &=
					f_{\chi}\.\frac{ \rho_{\rm eq} }{ 2 }\!
					\left(
						\frac{ M }{ \,M_{\odot} }
					\right)^{\!3/4}
					\left(
						\frac{ \bar r }{ r }
					\right)^{\!9/4}
					\quad
					[
						r
						<
						{ \bar r } ( M / M_{\odot} )^{1/3}
					]
					\, ,
\end{split}
\end{align}
where $\bar r \equiv \left(2\.G M_{\odot}\,t_{\rm eq}^{2} \right)^{1/3} \approx 0.0193\,{\rm pc}$ is the turn-around radius at $t_{\rm eq}$ for a solar-mass PBH and the energy density then is $\rho_{\rm eq} = 3\.H_{\rm eq}^{2}\,m_{\rm Pl}^{2} / ( 8 \pi )$. The factor $f_{\chi}$ ensures the WIMP density profile scales with the WIMP fraction and the mass in the WIMP halo at this stage is $f_{\chi}$ times the PBH mass.

For PBHs which form before KD, there is also a constant density core within $r_{\rm ta}( t_{\rm KD} )$, so for $M < M_{\rm KD}$ we can express the overall density profile at $t_{\rm eq}$ as
\begin{equation}
	\label{eq:WIMPprofile-ini}
	\tilde \rho_{i}( r )
		=
					\frac{ \rho_{\rm \chi, \,spike}( r ) \,\rho_{\rm KD} }
					{ \rho_{\rm \chi, \,spike}( r ) + \rho_{\rm KD} }
					\, .
\end{equation}
This expression reduces to $\rho_{\rm \chi, \,spike}( r )$ for $r \gg r_{\rm ta}(t_{\rm KD})$ and $\rho_{\rm KD}$ for $r \ll r_{\rm ta}(t_{\rm KD})$, with the transition occurring where $\rho_{\rm \chi, \,spike}( r ) \approx \rho_{\rm KD}$. This is similar to taking the halo profile to be $\rho_{i}( r ) = \min[\rho_{\rm \chi, \,spike}, \rho_{\rm KD}]$, as assumed by~\cite{Adamek:2019gns}. We have used the form in Eq.~\eqref{eq:WIMPprofile-ini} because it gives a smoother transition between the two expressions. For PBHs larger than $M_{\rm KD}$, which form after $t_{\rm KD}$, there is no constant-density core region and so Eq.~\eqref{eq:WIMPprofile-ini} is inapplicable. Instead, one applies Eq.~\eqref{eq:densityprofile} over the range $r_{\Srm} < r < r_{\rm ta}( t_{\rm eq})$. 

We refer to the expression given by Eq.~\eqref{eq:WIMPprofile-ini} as the ``initial'' density profile. However, it is not initial in the sense that it applies at a particular time: the density at a given value of $r$ within the halo is constant and it is only the halo radius which changes. It is initial in the sense that it neglects two physical effects which modify the profile. The relationship between these effects is rather obscure in previous literature, so we now clarify this. We present the full analysis in Appendix~\ref{sec:Sudden-Accretion} but here summarise the qualitative conclusions.

The first effect is that the WIMPs are expected to have a Maxwellian velocity distribution and the low-velocity ones will be captured by the black hole before those with the average velocity $\sigma$. This problem was studied by \cite{Eroshenko:2016yve}, who showed that Eq.~\eqref{eq:WIMPprofile-ini} is replaced with a more complicated expression, Eq.~\eqref{eq:rhoEroshenkoA},which also depends on the WIMP velocity distribution. Starting from $\tilde \rho_{i}$, this yields a new distribution $\rho_{i}$ and turns the $r^{-9/4}$ profile into an $r^{-3/2}$ profile at small $r$. However, a central core with constant density $\rho_{\rm KD}$ is retained since the density can never exceed this.

The second effect is that the above treatment does not take into account the orbital motion of WIMPs bound to the PBH (i.e.\ it neglects their kinetic energy). This problem was studied by~\cite{Adamek:2019gns} and concerns the mean velocity $\sigma$ rather than the low-velocity tail. (Here $\sigma$ applies at the moment of turn-around and not after virialisation.) Since the WIMP velocity decreases as $a^{-1}$, the kinetic and potential energies for WIMPs binding at a distance $r$ from the PBH at the time $t$ are
\begin{equation}
	E_{\Krm}
		=
					 k_{\Brm}\.T_{\rm KD}
					\left(
						\frac{ t_{\rm KD} }{ t }
					\right)
					,
	\quad
	E_{\prm}
		=
					\frac{ GM m_{\chi} }{ r }
		=
					\frac{m_{\chi} r_{\Srm} }{ 2\.r }
					\, ,
\end{equation}
where the first expression also specifies the velocity dispersion $\sigma$. Using Eq.~\eqref{eq:turnaround} to express $t$ in terms of $r$ then implies that the ratio of the two energies is
\begin{equation}
	\label{eq:energyratio}
	\frac{ E_{\Krm} }{ E_{\prm} }
		=
					\frac{ 2\.k_{\Brm}\.T_{\rm KD} }{ m_{\chi} }\.
					\frac{ t_{\rm KD} }{ \sqrt{r_{\Srm}\,r} }
					\, ,
\end{equation}
so the kinetic term is important out to a radius
\begin{align}
	\label{eq:rKdef}
	r_{\Krm}
		&\approx
					\frac{ t_{\rm KD}^{2} }{r_{\Srm}}
					\left(
						\frac{ 2\.k_{\Brm}\.T_{\rm KD} }{ m_{\chi} }
					\right)^{\!2} \\ \nonumber
		&\approx
					\left(
						\frac{ m_{\chi} }{ M_{\rm {Pl}} }
					\right)^{\!-9/2}
					\frac{ R_{\rm {Pl} }^{2} }{ r_{\Srm} }
		\approx
					10
					\left(
						\frac{ m_{\chi} }{ {\rm TeV} }
					\right)^{\!-9/2}
					\left(
						\frac{ M }{ M_{\odot} }
					\right)^{\!-1}
					{\rm cm}
					\, .
\end{align}
Outside this radius, we recover the spike solution of Eq.~\eqref{eq:densityprofile} except that the density must be multiplied by a concentration parameter $\alpha_{\Erm} \approx 1.53$, as shown by Eq.~\eqref{eq:rhoEroshenkocc}.
We note that
\begin{equation}
	r_{\rm ta} (t_{\rm eq} )
		\approx
					( r_{\Srm}\,t_{\rm eq}^{2} )^{1/3}
		\approx
					10^{9}\.( M / M_{\odot} )^{1/3}\,
					{\rm cm}
					\, ,
\end{equation}
and this exceeds $r_{\Krm}$ provided $M < 10^{12}\,M_{\odot} ( m_{\chi} / {\rm TeV} )^{-27/4}$. Thus one expects the $r^{-9/4}$ profile to develop well before the matter-dominated era.

These two effects are linked because the low-velocity tail is only important if $\sigma$ is sufficiently large. For example, there could be no such tail if $\sigma = 0$. Indeed, as shown in the Appendix, both effects produce an $r^{-3/2}$ tail within the radius $r_{\Krm}$. Finally, as we show in the Appendix, a profile with a $r^{-3/4}$ distribution develops at the core. The overall density profile is therefore 
\begin{equation}
	\label{eq:densityprofile2}
	\rho_{\rm \chi, \,spike}( r )
		=
					\begin{cases}
						f_{\chi}\.\rho_{\rm KD}\.
						\big(
							\frac{ r_{\Crm} }{ r }
						\big)^{3/4}
						,
							& \hbox{for $r \leq r_{\Crm}$}
						\, ,
						\\[2mm]
						f_{\chi}\.\frac{ \rho_{\rm eq} }{ 2 }\.
						\big(
							\frac{ M }{ \,M_{\odot} }
						\big)^{3/2}\.
						\big(
							\frac{ \hat r }{ r }
						\big)^{\!3/2}
							& \hbox{for $r_{\Crm} < r \leq r_{\Krm}$}
							\, ,
							\\[2mm]
						f_{\chi}\.\frac{ \rho_{\rm eq} }{ 2 }\.
						\big(
							\frac{ M }{ \,M_{\odot} }
						\big)^{3/4}\.
						\big(
							\frac{ \bar r }{ r }
						\big)^{9/4}
							& \hbox{for $r > r_{\Krm}$}
							\, .
							\\[-1mm]
\end{cases}
\end{equation}
Here $r_{\Crm}$ is the intersect of the first two lines, given by 
\begin{equation}
	r_{\Crm}
		\approx
						r_{\Srm}
						\left(
							\frac{ m_{\chi} }{ T_{\rm KD} }
						\right)
		\approx
						10^{9}
						\left(
							\frac{ M }{ M_{\odot} }
						\right)\!
						\left(
							\frac{ m_{\chi} }{ {\rm TeV} }
						\right)^{\!-1/4}
						{\rm cm} 
						\label{eq:C}
\end{equation}
and
\begin{equation}
	\hat r
		=
					G\.M_{\odot}\,
					\frac{ t_{\rm eq} }{ t_{\rm KD} }\,
					\frac{ m_{\chi} }{ T_{\rm KD} }
		=
					1.1 \times 10^{26}{\rm\,cm}\,
					( m_{\chi} / {\rm TeV} )^{2.25}
					\,,
\end{equation}
and the intersect of the second two lines is the radius $r_{\Krm}$ given by Eq.~\eqref{eq:rKdef}.

At KD, Eq.~\eqref{eq:energyratio} implies that the kinetic term can be neglected for~\citep{Adamek:2019gns}, 
\begin{equation}
	\label{eq:mass-range}
	M
		\gtrsim
					M_{\Krm}
		\approx
					10^{-5}\,M_{\odot}\.
					( m_{\chi} / {\rm TeV} )^{-17/8}
					\, .
\end{equation}
This is also the condition $r_{\Krm}$ exceeds $r_{\Crm}$, ensuring that the intermediate $r^{-3/2}$ region exists. Adamek {\it et al.} were mainly interested in solar-mass PBHs, so only needed to confirm that condition \eqref{eq:mass-range} is well satisfied. In this paper, we consider a much wider range of masses, including those for which the WIMP kinetic energy is important. Note that $M_{\Krm}$ is much less than $M_{\rm KD}$ for $m_{\chi} < 1$~TeV. They also discuss the effect of primordial power spectrum on the halo profile but conclude that this is not important.

\subsection{Halo formation after matter-radiation equality}

Although Eq.~\eqref{eq:turnaround} no longer applies after $t_{\rm eq}$, because the mass of the WIMP halo exceeds the mass of the black hole, the halo continues to grow. Indeed, self-similar secondary infall and virialisation should give a DM spike with the same radial dependence as Eq.~\eqref{eq:densityprofile}. This is confirmed by the numerical calculations in~\cite{Adamek:2019gns} and can be understood as follows. Since the black hole represents an initial overdensity $M / \tilde{M}$ for a region of mass $\tilde{M}$ and density fluctuations grow as $( 1 + z )^{-1}$ during the matter-dominated era, the mass gravitationally bound by the PBH grows as 
\begin{equation}
	\tilde{M}( z )
		=
					M\!
					\left(
						\frac{ 1 + z_{\rm eq} }
						{ 1 + z }
					\right)
					\label{eq:bound}
\end{equation}
after $t_{\rm eq}$. Since Eq.~\eqref{eq:bound} implies that the radius of the shell binding at redshift $z$ is $r \propto ( \tilde{M} / \rho )^{1/3} \propto ( 1 + z )^{-4/3} \propto \rho^{-4/9}$, this gives the same profile as Eq.~\eqref{eq:densityprofile} but it now extends beyond the radius $r_{\rm ta}( t_{\rm eq} )$.

\begin{figure}
	\vs{1mm}
	\includegraphics[width = 0.95\linewidth]{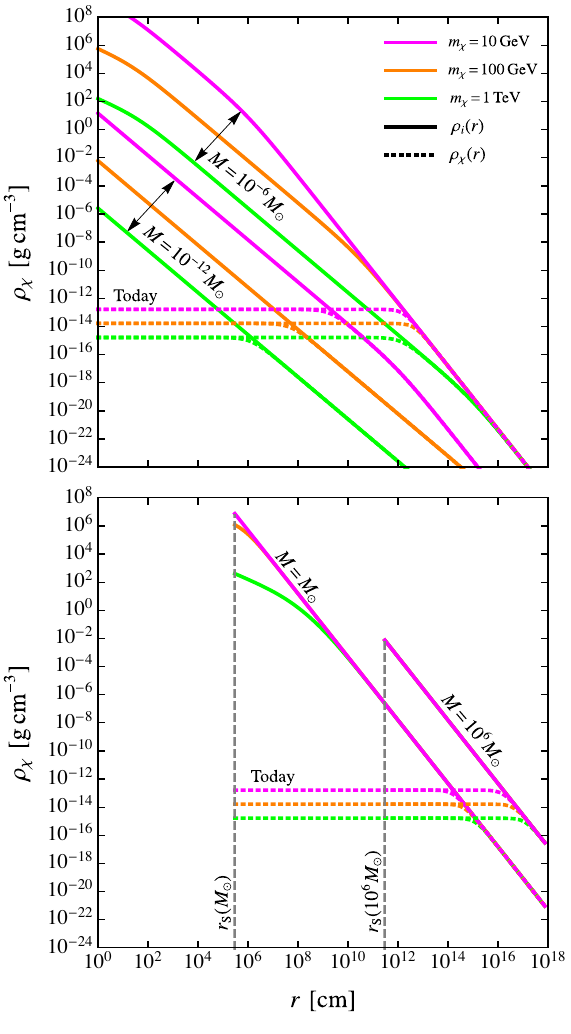} 
	\caption{Density profile of WIMPs bound to a PBH of mass $M = 10^{-12}\,M_{\odot}$ or $M = 10^{-6}\,M_{\odot}$
			 (top panel) and $M = 1\,M_{\odot}$ or $M = 10^{6}\,M_{\odot}$ (bottom panel) for $f_{\chi} \simeq 1$. 
			 We set 
			 $m_{\chi} = 10\,$GeV (magenta), 
			 $m_{\chi} = 100\,$GeV (orange) and 
			 $m_{\chi} = 1\,$TeV (green).
			 The density profiles before WIMP annihilations, $\rho_{i}( r )$, are shown by the solid lines and 
			 derived from Eq.~\eqref{eq:rhoEroshenkoA}. The density profiles after annihilations, 
			 $\rho_{\chi}( r )$, are shown by the dotted lines and labelled ``Today''. 
			 The plateau in the WIMP distribution is described by Eq.~\eqref{eq:rhomax} 
			 but does not apply for $r < r_{\Srm}$ (i.e.\ to the left of the vertical dashed lines 
			 in the lower diagram).}
	\label{fig:rhochi}
\end{figure}

So long as one neglects the effects of WIMP annihilations, the density profiles for different values of $M$ and $m_{\chi}$ are as indicated by the solid lines in Fig.~\ref{fig:rhochi}. These have been calculated numerically from Eq.~\eqref{eq:rhoEroshenkoA} but their qualitative form is as anticipated above. We have set $m_{\chi} = 10\,$GeV (magenta), $m_{\chi} = 100\,$GeV (orange) and $m_{\chi} = 1\,$TeV (green), this covering the most plausible range of values. The upper panel shows the profiles for $M = 10^{-12}\,M_{\odot}$ and $M = 10^{-6}\,M_{\odot}$, where one sees the transition from the constant-density region to the $r^{-3/2}$ region and then the $r^{-9/4}$ region. The lower panel shows the profiles for $M = 1\,M_{\odot}$ and $M = 10^{6}\,M_{\odot}$. In this case, there is no $r^{-3/2}$ region because $M > M_{\Krm}$ but there is still a constant-density region for $M < M_{\rm KD}$. The profiles are inapplicable inside the Schwarzschild radius but this is only relevant for the lower figure.

The formation of WIMP halos around PBHs due to adiabatic accretion at late times has also been studied by~\cite{Gondolo:1999ef}. If the WIMPs initially have a cusp profile scaling as $ r^{-\gamma}$, the presence of the black hole leads to a spike profile scaling as $r^{-\gamma_{\rm sp}}$, where $\gamma_{\rm sp} = ( 9 - 2\.\gamma ) / ( 4 - \gamma )$. This result was first derived in \cite{1995ApJ...440..554Q} and can also be derived from our Eq.~\eqref{eq:rhoEroshenkoA}. The DM distribution around the black hole is therefore steeper than in the surrounding cusp ($\gamma_{\rm sp} > \gamma$) providing $\gamma < 3$, as expected in most DM models, and the usual result ($\gamma_{\rm sp} = 9/4$) is obtained for a constant density profile ($\gamma = 0$).

This analysis no longer applies after the epoch of galaxy formation, which we take to be $z_{\star} \sim 10$, since the local density is no longer the background cosmological density. Astrophysical processes - in particular, tidal stripping - could modify the WIMP halos around BHs within galaxies. When a star passes near a BH, it deposits energy into the halo, which could remove part of it~\citep{Green:2006hh}. This mechanism has been invoked for self-gravitating halos made of WIMPs~\citep{Schneider:2010jr} or axions~\citep{Tinyakov:2015cgg, 2020arXiv201105377K} and it has recently been applied to WIMP halos around BHs~\citep{Hertzberg:2019exb}. Part of the halo could also be removed by the interaction amongst PBH-halo systems, particularly in high-density regions such as galactic centres or PBH clusters.

\subsection{Effect of WIMP Annihilations}

The WIMP population inside the halo is consumed by self-annihilation~\citep{Berezinsky:1992mx}. In order to estimate the density of WIMPs in the core of the distribution, we compare the inverse of the age of the halo $t_{\rm halo}$ with the self-annihilation rate $\Gamma_{\rm ann} = n_{\chi}\,\langle \sigma v \rangle^{}_{\Hrm}$. Here $\langle \sigma v \rangle^{}_{\Hrm}$ is the velocity-weighted cross-section in the halo, where the WIMPs are assumed to have a Boltzmann velocity distribution with dispersion $v_{\rm rms}$. Setting their density to be $\rho_{\chi} = m_{\chi}\.n_{\chi}$, the maximum WIMP concentration at redshift $z$ is
\begin{equation}
	\label{eq:rhomax}
	\rho_{\rm \chi, \,max}( z )
		 = 
					f_{\chi}\.
					\frac{ m_{\chi}\,H( z ) }
					{ \langle \sigma v \rangle^{}_{\Hrm} }
					\, ,
\end{equation}
where we have assumed $t_{\rm halo} \gg t_{\rm eq}$ and $t_{\rm halo} \sim 1 / H( z )$, where $H( z )$ is the Hubble rate at redshift $z$. Equation~\eqref{eq:rhomax} extends the result of previous literature~\citep{Ullio:2002pj, Scott:2009tu, Josan:2010vn} to an arbitrary redshift and WIMP fraction. For the Taylor expansion in Eq.~\eqref{eq:taylor}, the velocity-averaged cross-section leads to $\langle \sigma v \rangle^{}_{\Hrm} = a + 3\.b\.v_{\rm rms}^{2}$, so it generally differs from the thermal average $\langle \sigma v \rangle^{}_{\rm th}$ when higher-order terms in the expansion are taken into account. In the following, we neglect these terms in the expansion of Eq.~\eqref{eq:taylor} and set $\langle \sigma v \rangle^{}_{\Hrm} = \langle \sigma v \rangle^{}_{\rm th}$.

The WIMP profile is then
\begin{equation}
	\label{eq:WIMPprofile}
	\rho_{\chi}
		=
					\frac{ \rho_{i}( r )\,\rho_{\rm \chi, \,max}( z ) }
					{ \rho_{i}( r ) + \rho_{\rm \chi, \,max}( z ) }
					\, ,
\end{equation}
with the plateau in Eq.~\eqref{eq:rhomax} extending to the radius $r_{\rm cut}$, which from Eq.~\eqref{eq:rhoEroshenkoA} is defined implicitly by
\begin{equation}
	\label{eq:definercut}
	\tilde \rho_{i}( r_{\rm cut} )
		\approx
					\rho_{\rm \chi, \,max}( z )
					\, .
\end{equation}
Even though both $\tilde \rho_{i}( r )$ and $\rho_{\rm \chi, \,max}( z )$ are proportional to $f_{\chi}$, so these dependencies cancel out in Eq.~\eqref{eq:definercut}, $r_{\rm cut}$ still depends on $f_{\chi}$ through the cross-section $\langle \sigma v \rangle^{}_{\Hrm}$. For example, if the WIMP kinetic energy can be neglected, we obtain
\begin{equation}
	\label{eq:rcut}
	\frac{ r_{\rm cut} }{ \bar r }
		=
					\!\left[
						\alpha_{\Erm}\,\frac{ \rho_{\rm eq} }{ 2 }\!
						\left(
							\frac{ M }{ \,M_{\odot} }
						\right)^{\!3/4}
						\frac{ \langle \sigma v \rangle^{}_{\Hrm} }
						{ m_{\chi}\,H( z ) }
					\right]^{4/9}
					.
\end{equation}
The lines labelled ``Today'' in Fig.~\ref{fig:rhochi} show the density profiles implied by Eq.~\eqref{eq:WIMPprofile} for the various values of $M$ and $m_{\chi}$ when we fix $f_{\chi} = 1$ and $z = 0$. The profiles are characterised by a plateau region for the inner orbits, given by Eq.~\eqref{eq:rhomax} and shown by dotted lines. In the outer region, the profile is described by the solution to Eq.~\eqref{eq:rhoEroshenkoA}.

We now determine the WIMP annihilation rate around each black hole. This proceeds at the rate
\begin{equation}
	\label{eq:decayrate-gal}
	\Gamma_{0}
		=
					\frac{ \langle \sigma v \rangle^{}_{\Hrm} }
					{ m_{\chi}^{2} }\,
					\int\d V\.\rho_{\chi}^{2}
					\, ,
\end{equation}
where the integration is taken over the volume of the WIMP halo. In some sense, each black hole resembles a decaying particle of mass $M$ and decay rate $\Gamma_{0}$.

We assume that the WIMP density profile around a PBH is described by Eq.~\eqref{eq:rhomax} in the inner part but falls off as $r^{-\alpha}$ with $\alpha > 0$ in the outer part. Then the decay rate~\eqref{eq:decayrate-gal} at the current time $t_{0}$ becomes
\begin{align}
\begin{split}
	\label{eq:Gamma-alpha}
	\Gamma_{0}
		&=
					\frac{ 4\pi\.\alpha\,f_{\chi}^{2}\,H_{0} \rho_{\rm eq} }
					{ 3\.(2 \alpha - 3 )\.m_{\chi} }\!
					\(
						\frac{ \langle \sigma v \rangle^{}_{\Hrm}\,\rho_{\rm eq} }
						{ 2\.m_{\chi}\,H_{0} }
					\)^{\!3/\alpha-1}
					r^{3}_{\rm ta}( t_{0} )
					\\[1.5mm]
		&=
					\!\left[
						\frac{ 8\pi\.G t_{0}^{2}\,\alpha\,H_{0}\rho_{\rm eq} }
						{ 3\.(2 \alpha - 3 )\.m_{\chi} }\!
						\(
							\frac{\langle \sigma v \rangle^{}_{\rm DM}\,\rho_{\rm eq} }
							{ 2\.m_{\chi}\,H_{0} } 
						\)^{\!3/\alpha - 1}
					\right]
					M\,f_{\chi}^{3 - 3/\alpha}
					\, ,
\end{split}
\end{align}
where in the last step we have used the expression for $\langle \sigma v \rangle^{}_{\Hrm}$ in Eq.~\eqref{eq:sigmavth} and the definition for $r_{\rm ta}( t )$ in Eq.~\eqref{eq:turnaround}. When the WIMP kinetic energy can be neglected, $\alpha = 9/4$ and Eq.~\eqref{eq:decayrate-gal} gives
\begin{equation}
	\label{eq:Gamma}
	\Gamma_{0}
		=
					\frac{ 3 }{ 8 }
					\left(
						\frac{ \langle \sigma v \rangle^{}_{\Hrm}\,
						\rho_{\rm eq}\,H_{0}^{2} }
						{ 2\.m_{\chi}^{4} }
					\right)^{\!1/3}
					f_{\chi}^{2}\,M
		\equiv
					\Upsilon\.
					f_{\chi}^{1.7}\,
					\frac{ M }{ M_{\odot} }
					\, ,
\end{equation}
where the quantity $\Upsilon$ has units of ${\rm s^{-1}}$. Numerically,
\begin{equation}
	\label{eq:upsilon}
	\Upsilon
		=
					1.2\times 10^{34}{\rm\, s^{-1}}\,
					( {\rm TeV} / m_{\chi} )^{4/3}
					\,.
\end{equation}
When the WIMP kinetic energy is important, Eq.~\eqref{eq:decayrate-gal} with the profile~\eqref{eq:densityprofile2} gives
\begin{equation}
	\label{eq-Gamma0Kin}
	\Gamma_{0}
		=
					\frac{ \langle \sigma v \rangle^{}_{\Hrm} }
					{ m_{\chi}^{2} }\,\frac{ \pi \rho_{\rm eq}^{2} }{ 3 }\!
					\left(\!
						G\frac{t_{\rm eq}}{t_{\rm KD}}
						\frac{m_{\chi}}{T_{\rm KD}}\!
					\right)^{\!3}\!
					f_{\chi}^{2}\.M^{3}
		\equiv
					\Theta f_{\chi}
					\left(
						\frac{ M }{ M_{\odot} }
					\right)^{\!3}
					\, ,
\end{equation}
where the quantity $\Theta$ has units of ${\rm s^{-1}}$. Numerically,
\begin{equation}
	\label{eq:Theta}
	\Theta
		=
					3.2 \times 10^{57}
					{\rm\, s^{-1}}\,(m_{\chi}/{\rm TeV} )^{4.75}
					\, .
\end{equation}

\section{Flux of gamma-rays from WIMP Annihilation}
\label{sec:Gamma--Ray-Flux-from-WIMP-Annihilation}

The usual assumption in previous analyses is that WIMPs provide most of the dark matter ($f_{\chi} \approx 1$), this then implying strong constraints on $f_{\rm PBH}( M )$. Indeed, we follow this approach in Sections \ref{sec:Galactic-Background-Flux} to \ref{sec:Combined-Results} below. However, motivated by the current interest in PBHs, we also consider the possibility that PBHs comprise most of the DM, this then placing interesting constraints on $f_{\chi}$. In the intermediate situation, in which $f_{\rm PBH}( M )$ is significant but less than $1$, we will conclude that one needs a third DM component.

\subsection{Galactic Background Flux}
\label{sec:Galactic-Background-Flux}

We assume that the distribution of PBHs in the Milky Way tracks the distribution of DM in the halo $\rho_{\Hrm}( R )$, scaled by the fraction $f_{\rm PBH}$, where $R$ is the Galactocentric distance. The expected flux (s$^{-1}$\,cm$^{-2}$) of $\gamma$-rays from the annihilation of WIMPs bound to PBHs is then~\citep{Ullio:2002pj}
\begin{equation}
	\label{eq:galflux}
	\Phi_{\gamma, {\rm \,Gal}}
		=
					\frac{ f_{\rm PBH}\.\Gamma_{0} }{ M }\.
					N_{\gamma}\.D( b, \ell )
					\, .
\end{equation}
Here $N_{\gamma}$ is the number of detectable photons resulting from annihilations:
\begin{equation}
	\label{eq:Ngamma}
					N_{\gamma}( m_{\chi} )
		= 
					\int_{E_{\rm th}}^{m_{\chi}}\!\d E\;
					\frac{ \d N_{\gamma} }{ \d E }
		\approx
					18\,( m_{\chi} / {\rm TeV} )^{0.3}
					\, ,
\end{equation}
where $\d N_{\gamma} / \d E$ is the spectrum of $\gamma$-rays from each halo~\citep{Cirelli:2009dv} and $E_{\rm th}$ is the threshold energy for detection. Numerical expressions for the spectrum come from the code of~\cite{Cirelli:2010xx}, see also~\cite{Amoroso:2018qga}, and the last expression in Eq.~\eqref{eq:Ngamma} is a fit to the numerical solution for different values of $m_{\chi}$.

In Eq.~\eqref{eq:galflux}, the factor $D$ is the integral over the solid angle of the telescope's field of view $\Omega$ and along the line of sight (los) in the direction with Galactic coordinates $( b, \ell )$. Since we are at a distance $R_{\odot} \approx 8.5\,$kpc from the Galactic centre, this is
\begin{equation}
	D( b, \ell )
		=
					\frac{ \Omega }{ 4\pi }\.
					\int_{\rm los}\d s\;
					\rho_{\Hrm}( R )
					\, 
\end{equation}
where the Galactocentric distance $R = R\.( s, b, \ell )$ is related to the los distance $s$ by
\begin{equation}
	R\.( s, b, \ell )
		=
					\sqrt{s^{2} + R_{\odot}^{2} - 2 R_{\odot}\.s\.\cos b\.\cos\ell\,}
					\, .
\end{equation}
The dependence of the flux in Eq.~\eqref{eq:galflux} on the direction of observation $( b, \ell )$ has been incorporated, for example, by~\cite{Carr:2016hva} in computing the sky map for the Hawking emission of PBHs in the Galactic halo. Here we focus on the direction of the Galactic centre, where we expect the flux to be strongest.

In principle, it is possible to detect $\gamma$-ray `point' sources associated with WIMP annihilations around the PBHs, as shown in the detailed analysis of~\cite{Bertone:2019vsk}. Here we consider the detectability of the diffuse $\gamma$-ray background from WIMP annihilations by the Fermi Large Area Telescope (LAT)~\citep{2010PhRvL.104j1101A}.\footnote{While Fermi currently offers the best sensitivity for the WIMP mass range considered here, the Cherenkov Telescope Array~\citep{CTAConsortium:2018tzg} and Large High Altitude Air Shower Observatory~\citep{Bai:2019khm} are more sensitive above a TeV.} The signal can be analysed with a likelihood analysis in which each Fermi-LAT energy bin is compared with the sum of the fluxes from WIMP annihilations and other astrophysical sources, a method used by~\cite{Ackermann:2015tah, DiMauro:2015tfa, Ando:2015qda}. We take the upper limit to the WIMP flux to be $\Phi_{\rm res} \approx 10^{-7}{\rm \,cm^{-2}\,s^{-1}}$, corresponding to the residual component of the Fermi $\gamma$-ray flux after other astrophysical sources are subtracted. This is about one order of magnitude larger than the Fermi point-source sensitivity, $\Phi_{\rm Fermi} = 6 \times 10^{-9}\,{\rm \,cm^{-2}\,s^{-1}}$.

The condition $\Phi_{\gamma,\,{\rm gal}} \leq \Phi_{\rm res}$ yields
\begin{align}
	\label{eq:galbound}
	f_{\rm PBH}
		&\lesssim
					\frac{ \Phi_{\rm res}\.M }
					{ D( b, \ell )\.\Gamma_{0}\.N_{\gamma} }
					\\[1.5mm]
		&\approx
					\begin{cases}
						8.7\times 10^{-8}
						\left(
							\frac{ m_{\chi} }{ \rm TeV }
						\right)^{\!1.0}
							& \hbox{($M\gtrsim M_{*}$)}
						\\[1mm]
						3.3 \times 10^{-11}
						\left( \frac{ m_{\chi} }{ \rm TeV } \right)^{\!-5.07}\!
						\left(\!
							\frac{ 10^{-10}\.M_{\odot} }{ M }
						\!\right)^{\!2}
							& \hbox{($M \lesssim M_{*}$)}
					\end{cases}
					\, ,
					\nonumber
\end{align}
where $M_{*}$ is the intersect of the last two expressions,
\begin{equation}
	\label{eq:masschange}
	M_{*}
		\approx
					2 \times 10^{-12}\,M_{\odot}\,
					( m_{\chi}/{\rm TeV} )^{-3.0}
					\, .
\end{equation}
The first expression applies when the WIMP kinetic energy can be neglected and is derived analytically from Eq.~\eqref{eq:Gamma}. The second condition includes the effect of the WIMP kinetic energy and comes from the result given in Eq.~\eqref{eq-Gamma0Kin}. Note that $M_{*}$ is considerably less than the mass~\eqref{eq:mass-range} where kinetic energy can be {\it completely} neglected. This is because the transition is only gradual and $M_{*}$ just corresponds to the intersect of the asymptotic expressions.

{The $f_{\rm PBH}$ constraint given by the flat part of Eq.~\eqref{eq:galbound} is indicated for various WIMP masses in Table~\ref{tab:tab1}, where we assume $f_{\chi} \approx 1$. The full constraints are shown by the red lines in Fig.~\ref{fig:fPBH} for $2 \times 10^{-12}\.m^{-3.2} \lesssim M / M_{\odot} \lesssim 8 \times 10^{4}\.m^{1.0}$, where $m \equiv m_{\chi}$/TeV. The different curves correspond to a WIMP mass of $10\,$GeV (dashed line), $100\,$GeV (solid line) and $1\,$TeV (dotted line). The Galactic population of PBHs can also be bound from below by requiring that there be at least one of them within our halo~\citep{Carr:1997cn}. This gives
\begin{equation}
	\label{eq:IL}
	f_{\rm PBH}
		\gtrsim
					\frac{ M }{ M_{\Erm} }
					\, ,
\end{equation}
where $M_{\Erm} \approx 10^{12}\,M_{\odot}$ is the total halo mass. This intersects the upper bound from WIMP annihilations at a mass 
\begin{equation}
	\label{eq:ILgal}
	M^{\rm gal}
		=
					M_{\Erm}\,
					\frac{ M_{\odot}\.\Phi_{\rm res} }
					{ \Upsilon\.N_{\gamma}\.D( b, \ell ) }
		\approx
					8 \times 10^{4}\,M_{\odot}\,
					( m_{\chi} / {\rm TeV} )^{1.0}
					\, .
\end{equation}
The last exponent is derived using $\Upsilon \propto m_{\chi}^{-4/3}$ and the numerical fit for $N_{\gamma}( m_{\chi} )$.

\begin{table}
\def\arraystretch{1.5}
	\begin{tabular}{ccccc}
	\hline\hline
		\;
		$m_{\chi}\,$(TeV)	& $f_{\rm PBH}^{\rm Gal}$ 
							& $f_{\chi}^{\rm Gal}$ 
							& $f_{\rm PBH}^{\rm eg}$ 
							& $f_{\chi}^{\rm eg}$\;\\
		\hline
		$10^{-2}$			& $8 \times 10^{-10}$\;	
							& $3 \times 10^{-6}$	
							& $2 \times 10^{-11}$ 
							& $2 \times 10^{-7}$\\
		$10^{-1}$			& $8 \times 10^{-9}$\;	
							& $2 \times 10^{-5}$	
							& $2 \times 10^{-10}$ 
							& $1 \times 10^{-6}$\\
		$10^{0}$			& $8 \times 10^{-8}$\;	
							& $5 \times 10^{-5}$	
							& $2 \times 10^{-9}$ 
							& $5 \times 10^{-6}$\\
		$10^{1}$			& $9 \times 10^{-7}$\;	
							& $2 \times 10^{-4}$	
							& $3 \times 10^{-8}$ 
							& $2 \times 10^{-5}$\\
		\hline
	\end{tabular}
	\caption{Bounds from the Galactic (Gal) and extragalactic (eg) $\gamma$-ray flux on $f_{\rm PBH}$ 
		when the DM is mainly WIMPs and on $f_{\chi}$ when it is mainly PBHs
		for different WIMP masses. See main text for additional discussion.}
	\label{tab:tab1}
\end{table}

\begin{figure}
	\includegraphics[width = 0.99\linewidth]{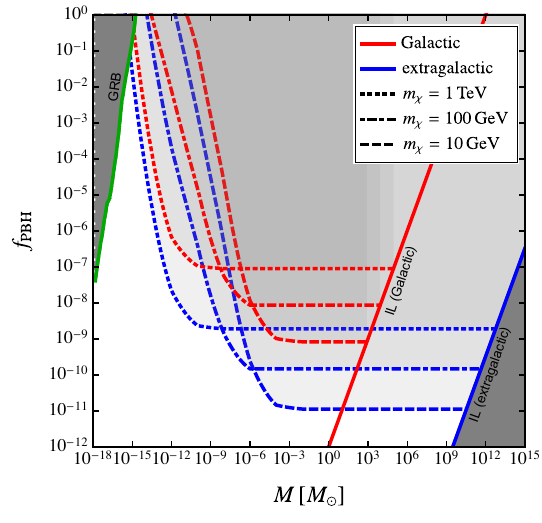} 
	\caption{Constraints on $f_{\rm PBH}$ as a function of PBH mass 
			from Galactic (red) or extragalactic (blue) $\gamma$-ray background.
 			Results are shown for 
			$m_{\chi} = 10\,{\rm GeV}$ (dashed lines), 
			$m_{\chi} = 100\,{\rm GeV}$ (dot-dashed lines) and 
			$m_{\chi} = 1\,{\rm TeV}$ (dotted lines), 
			setting $\sv = 3 \times 10^{-26}\,$cm$^{3}$/s.
			Also shown are the Galactic (red solid line) 
			and the extragalactic incredulity limits (blue solid line).}
	\label{fig:fPBH}
\end{figure}

\subsection{Extragalactic Background Flux}
\label{sec:Extragalactic-Background-Flux}

In order to discuss the extragalactic component, we adopt the standard flat $\Lambda$CDM cosmological model with current radiation density $\Omega_{\rrm} = 7 \times 10^{-5}$, matter density $\Omega_{\mrm} = 0.31$ and dark energy density $\Omega_{\Lambda} \approx 0.69$ in units of the critical density~\citep{Aghanim:2018eyx}. In contrast to other works~\citep{Handley:2019tkm, DiValentino:2019qzk, Vagnozzi:2020zrh}, we do not include the effects of cosmological curvature or the possibility that the dark energy evolves with redshift~\citep{Poulin:2018cxd, DiValentino:2019exe}. The Hubble rate at redshift $z$ is then $H( z ) = H_{0}\,h( z )$ with
\begin{align}
	h( z )
		&=
					\sqrt{
						\Omega_{\Lambda}
						+
						\Omega_{\mrm}\.( 1 + z )^{3}
						+
						\Omega_{\rrm }\.( 1 + z )^{4}
					\,}
					\, .
\end{align}
The extragalactic differential flux (cm$^{-2}$\,s$^{-1}$\,MeV$^{-1}$\,ster$^{-1}$) of $\gamma$-rays is produced by the {\it collective} annihilations of WIMPs around PBHs at all redshifts~\citep{Ullio:2002pj},
\begin{equation}
	\label{eq:flux}
	\frac{ \d\Phi_{\gamma} }{ \d E\.\d\Omega }\bigg|_{\rm eg}
		 \!\!\!=
		 			\int\limits_{0}^{\,\infty}\d z\.
					\frac{ e^{-\tau_{\Erm}(z,\.E)} }{ 8\pi H( z ) }
					\frac{ \d N_{\gamma} }{ \d E }
					\int\!\d M\;
					\Gamma( z )\.
					\frac{ \d n_{\rm PBH}( M ) }{ \d M }
					\, ,
\end{equation}
where ``eg'' indicates extragalactic, $n_{\rm PBH}$ is the number density of PBHs, $\Gamma( z )$ is the WIMP annihilation rate around each PBH, and $\tau_{\Erm}$ is the optical depth at redshift $z$ resulting from
	{\it (i)} photon-matter pair production,
	{\it (ii)} photon-photon scattering, and
	{\it (iii)} photon-photon pair production~\citep{Cirelli:2009dv, Slatyer:2009yq}. 
The numerical expressions for both the energy spectrum $\d N_{\gamma} / \d E$ and the optical depth are taken from~\cite{Cirelli:2010xx}.

When the WIMP velocity distribution can be neglected, the $z$-dependence of the decay rate obtained from Eq.~\eqref{eq:Gamma} becomes $\Gamma( z ) = \Gamma_{0}\,[ h( z ) ]^{2/3}$, where $\Gamma_{0} = \Upsilon\.f_{\chi}^{1.7}\,M / M_{\odot}$. We can then implement the normalisation of the PBH mass function,
\begin{equation}
	\label{eq:MF-normalise}
	\int\!\d M\;M\.\frac{ \d n_{\rm PBH}( M,\.z ) }{ \d M }
		\equiv 
					\rho_{\rm PBH}( z )
		=
					f_{\rm PBH}\,\rho_{\rm DM}( z )
					\, ,
\end{equation}
to integrate over the mass dependence in Eq.~\eqref{eq:flux}. Integrating over the energy and angular dependences leads to an expression for the flux
\begin{equation}
	\label{eq:flux-normalised}
	\Phi_{\gamma, {\rm \,eg}}
		 = 
					\frac{ f_{\rm PBH}\,
					\rho_{\rm DM} }{ 2 H_{0}\.M_{\odot} }\,
					\Upsilon\.f_{\chi}^{1.7}
					\tilde{N}_{\gamma}( m_{\chi} )
					\, ,
\end{equation}
where $\rho_{\rm DM}$ is the present dark matter density and $\tilde{N}_{\gamma}$ is the number of photons produced: 
\begin{equation}
	\label{eq:tildeNgamma}
	\tilde{N}_{\gamma}( m_{\chi} )
		\equiv
					\int_{z_{\star}}^{\infty}\!\d z\;
					\int_{E_{\rm th}}^{m_{\chi}}\!\d E\;
					\frac{ \d N_{\gamma} }{ \d E }
					\frac{ e^{-\tau_{\Erm}( z,\.E )} }
					{ [ h( z ) ]^{1/3} }
					\, .
\end{equation}
Here the lower limit in the redshift integral corresponds to the epoch of galaxy formation. We assume $z_{\star} \sim 10$ but changing it from $10$ to $15$ only leads to a $5\%$ decrease in the value of $\tilde{N}_{\gamma}( m_{\chi} )$, which is much smaller than the uncertainty from other sources.

The analysis becomes more complicated after $z_{\star}$. In particular, if the PBHs are small enough to be inside galaxies, then the growth of the WIMP halos is no longer determined by the background cosmological WIMP density. The halos may also be modified or even disrupted by various dynamical effects. We assume $z_{\star} \sim 10$ but changing it from $10$ to $15$ only leads to a $5\%$ decrease in the value of $\tilde{N}_{\gamma}( m_{\chi} )$, independent of the WIMP mass, which is much smaller than the uncertainty from other sources. Of course, if the PBHs are too large to be inside galaxies, then the epoch of galaxy formation is irrelevant.

Comparing the integrated flux with the Fermi sensitivity $\Phi_{\rm res}$ yields
\begin{align}
	\label{eq:egbound}
	f_{\rm PBH}
		&\lesssim
					\frac{ 2 M\,H_{0}\,\Phi_{\rm res} }
					{ \rho_{\rm DM}\,
					\Gamma_{0}\,\tilde{N}_{\gamma}( m_{\chi} ) }
					\\[1.5mm]
		&\approx
					\begin{cases}
						2 \times 10^{-9}\,
						( m_{\chi} / {\rm TeV} )^{1.1}
							& \hbox{($M \gtrsim M_{*}$)}
							\\[1.5mm]
						\!1.1\times 10^{-12}\left(
							\frac{ m_{\chi} }{\rm TeV}
						\right)^{-5.0}
						\left(
							\frac{ M }{ 10^{-10}\,M_{\odot} }
						\right)^{\!-2}
							& \hbox{($M \lesssim M_{*}$)}
					\end{cases}
					\, ,
					\nonumber
\end{align}
where $M_{*}$ is given by Eq.~\eqref{eq:masschange}. The numerical bounds for the flat part of this constraint are shown in Table~\ref{tab:tab1}. The full constraint is shown by the blue curves in Fig.~\ref{fig:fPBH} for a WIMP mass of $10\,$GeV (dashed line), $100\,$GeV (solid line) and $1\,$TeV (dotted line). We note that the extragalactic bound intersects the cosmological incredulity limit~\eqref{eq:IL} at a mass
\begin{equation}
	\label{eq:MiL}
	M_{\rm eg}
		=
					\frac{ 2\.H_{0}\.M_{\odot}\.\Phi_{\rm res}\. M_{\Erm} }
					{ \alpha_{\Erm}\.\rho_{\rm DM}\,
					\Upsilon\,\tilde{N}_{\gamma}( m_{\chi} ) }
		\approx
					5 \times 10^{12}\,M_{\odot}\,
					( m_{\chi}/{\rm TeV} )^{1.1}
					\, ,
\end{equation}
where we have used our fit for $\tilde N_{\gamma}( m_{\chi} )$ and set $M_{\Erm} \approx \rho_{\rm DM} / H_{0}^{3} \approx 3 \times 10^{21}\,M_{\odot}$.

\subsection{Combined Results}
\label{sec:Combined-Results}

We now comment on the $\fPBH$ constraints shown in Fig.~\ref{fig:fPBH}, these applying only if WIMPs provide most of the DM.\\[1mm]

\noindent
(1) The grey region at the top left of Fig.~\ref{fig:fPBH}, labelled ``GRB'', gives the current constraint on $f_{\rm PBH}$ from the soft $\gamma$-ray background generated by PBH evaporations~\citep{Carr:2009jm, Carr:2020gox, Coogan:2020tuf}. It is interesting that WIMP annihilations also give an ``effective'' black hole decay limit~\citep{Adamek:2019gns}, so both limits can be interpreted as being due to decays.
\\[1mm]

\noindent
(2) The extragalactic bound is always more stringent than the Galactic one, the ratio being
\begin{equation}
\label{eq:CIL}
	\frac{ f_{\rm PBH}^{\rm eg} }{ f_{\rm PBH}^{\rm gal} }
		\sim
					H_{0}\.r_{\odot}\.\Bigg(
						\frac{ N_{\gamma}( m_{\chi} ) }
						{ \tilde{N}_{\gamma}( m_{\chi} ) }
					\Bigg)\!
					\Bigg(
						\frac{ \rho_{\Hrm}(r_{\odot}) }
						{ \rho_{\rm DM}\. }
					\Bigg)
		\sim
					\Ocal\!\( 10^{-2} \)
					.
\end{equation}
Our argument cannot place a bound on the PBH fraction above the mass given by Eq.~\eqref{eq:IL} with $M_{\Erm} = 10^{12}\,M_{\odot}$ (red solid line) in the Galactic case or by Eq.~\eqref{eq:CIL} with $M_{\Erm} = 3 \times 10^{21}\,M_{\odot}$ (blue solid line) in the extragalactic case. Black holes above these bounds are not expected to populate the Galaxy or the Universe.
\\[1mm] 

\noindent
(3) We have included the effect of the WIMPs' initial velocity distribution in computing their density profiles. This is important below the PBH mass indicated by Eq.~\eqref{eq:mass-range}, which corresponds to the sloping curves in Fig.~\ref{fig:fPBH}. In this case, Eq.~\eqref{eq:rhoEroshenkoA} gives the halo profile, whereas the profile before contraction is given by Eq.~\eqref{eq:WIMPprofile-ini} in the higher mass range. The WIMP profile at redshift $z$ is then computed from Eq.~\eqref{eq:WIMPprofile}. We note that the sloping parts of the curves in Fig.~\ref{fig:fPBH} have been derived by~\cite{Eroshenko:2016yve} and the flat parts by~\cite{Adamek:2019gns}. However, this is the first analysis to cover the full PBH mass range.

\subsection{Constraints on the WIMP population}
\label{sec:Constraints-on-the-WIMP-population}

We now extend the above analysis to the case in which WIMPs do not provide most of the DM. The abundance of thermally-produced WIMPs is set at the onset of their chemical decoupling from the plasma, as discussed in Sec.~\ref{sec:Thermal-Production-of-WIMPs}. With the thermal freeze-out mechanism, the WIMP abundance is determined by properties such as the WIMP mass and its interactions within the SM. Although there are currently no bounds on $f_{\chi}$ in the mass range $m_{\chi} \gtrsim 1\,$GeV, this parameter affects the detection of $\gamma$-rays from WIMP annihilations~\citep{Duda:2001ae, Baum:2016oow}.

We first modify the above analysis to place a bound on WIMPs if $f_{\rm PBH} + f_{\chi} = 1$ but with PBHs providing most of the DM. Since the extragalactic flux~\eqref{eq:flux-normalised} is still bound by the sensitivity $\Phi_{\rm res}$, we can proceed as in Sec.~\ref{sec:Extragalactic-Background-Flux} but considering the solution with $f_{\chi} \ll f_{\rm PBH}$. The decay rate is given by Eq.~\eqref{eq:Gamma-alpha} and this leads to the extragalactic bound
\begin{equation}
	f_{\chi}
		\lesssim
					\left(
						\frac{ 2 M\.H_{0}\,\Phi_{\rm res} }
						{ \rho_{\rm DM}\,\Gamma_{0}\,\tilde{N}_{\gamma}( m_{\chi} ) }
					\right)^{\!0.6}
\end{equation}
when the WIMP kinetic energy can be neglected. For different values of the WIMP mass, this gives the bounds shown in the fifth column of Table~\ref{tab:tab1}. A numerical fit in this case gives $f_{\chi} \lesssim 5.5 \times 10^{-5}\,( m_{\chi} / {\rm TeV} )^{0.6}$ and $5 \times 10^{-6}\,( m_{\chi} / {\rm TeV} )^{0.7}$ for the for the Galactic and extragalactic components, respectively.

Results are shown in Fig.~\ref{fig:fchi} with the values of $f_{\chi}$ indicated by the coloured scale as a function of $M$ (horizontal axis) and $m_{\chi}$ (vertical axis). The colour shows the maximum WIMP DM fraction if most of the DM comprises PBHs of a certain mass and complements the constraints of the PBH DM fraction if most of the DM comprises WIMPs with a certain mass and annihilation cross-section. In this situation, Fig.~\ref{fig:fPBH} can also be applied but all the constraints weaken as $f_{\chi}^{-1.7}$ from Eq.~\eqref{eq:Gamma}.

Clearly, the assumption $f_{\rm PBH} \approx 1$ used to derive the bound on $f_{\chi}$ cannot be applied for PBH mass ranges in which strong constraints on $f_{\rm PBH}$ can already be placed by other arguments. Indeed, there are only a few mass ranges in which one could have $f_{\rm PBH} \approx 1$. For example, this is still possible in the range $10^{-15} \text{--} 10^{-10}\,M_{\odot}$. In this case, depending on the value of $m_{\chi}$, the WIMP abundance could vary widely and even be close to $1$. However, if PBHs in the mass range $1\text{--}10\,M_{\odot}$ provide most of the DM, as argued by~\citet{Carr:2019kxo}, then one would require $f_{\chi} \lesssim 10^{-5}$ for all the WIMP masses considered. One could also consider a model with an extended PBH mass function, with a massive population attracting the WIMP halos and a lighter population providing most of the DM. This would still be compatible with the WIMP constraint since the limit on $f_{\rm PBH}( M )$ is weaker below the mass $M_{*}$, given by Eq.~\eqref{eq:masschange}, and independent of $M$ above this.

The important point is that even a small value of $f_{\rm PBH}$ may imply a strong upper limit on $f_{\chi}$. For example, if $M_{\rm PBH} \gtrsim 10^{-11}\,M_{\odot}$ and $m_{\chi} \lesssim 100\,$GeV, both the WIMP and PBH fractions are $\Ocal( 10\% )$. Since neither WIMPs nor PBHs can provide all the DM in this situation, this motivates us to consider situations in which $f_{\rm PBH} + f_{\chi} \ll 1$, requiring the existence of a third DM candidate (i.e.\ the ``something else'' of our title). Particles which are not produced through the mechanisms discussed above or which avoid annihilation include axion-like particles~\citep{Abbott:1982af, Dine:1982ah, Preskill:1982cy}, sterile neutrinos~\citep{Dodelson:1993je, Shi:1998km}, ultra-light or ``fuzzy'' DM~\citep{Hu:2000ke, Schive:2014dra}. Other forms of MACHOs could also serve this purpose.

\begin{figure}
	\includegraphics[width = 0.99\linewidth]{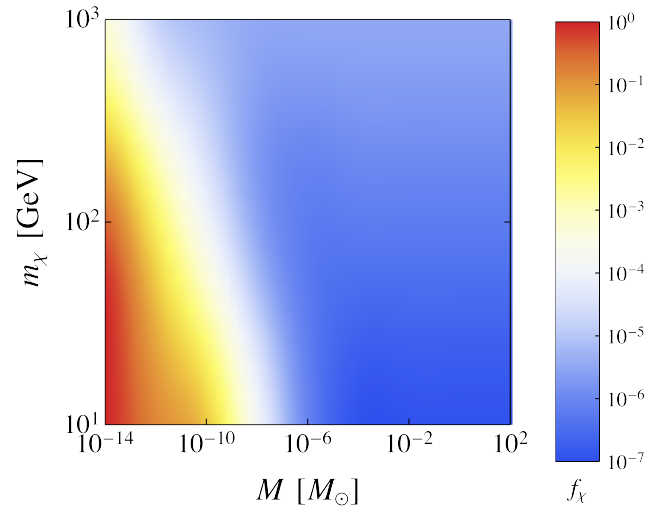} 
	\caption{The density plot shows the fraction of WIMPs $f_{\chi}$ (colour bar) 
			as a function of the PBH mass $M$ (horizontal axis) and 
			of the WIMP mass $m_{\chi}$ (vertical axis). 
			We fixed $f_{\rm PBH} + f_{\chi} = 1$.
			\vs{2mm}
			}
	\label{fig:fchi}
\end{figure}

\section{Is Something Else Implied by PBH Detections?}
\label{sec:Implications-from-Detections}

We now briefly review several observational hints that PBHs may exist. Each of these observations implies a lower limit on $f_{\rm PBH}( M )$ for some value of $M$ and the above argument then implies an upper limit on $f_{\chi}$ well below $1$. As indicated above, this suggests the existence of a third DM component.
\\[2mm]
{\it LIGO/Virgo Results{\;---\;}}The recent discovery of intermediate mass black hole mergers by the LIGO/Virgo collaboration~\citep{PhysRevLett.125.101102} might be the first direct detection of PBHs. It is unclear that these gravitational-wave events are primordial in origin, although it has been claimed that at least some fraction must be~\citep{Franciolini:2021tla}. However, if they are, the PBH DM fraction must be larger than $10^{-3}$. There might even be evidence for sub-solar candidates, which could only be primordial~\citep{Phukon:2021cus}. \\[2mm]
{\it Planetary-Mass Microlenses{\;---\;}}Using data from the five-year OGLE survey of 2622 microlensing events in the Galactic bulge~\citep{2017Natur.548..183M},~\cite{Niikura:2019kqi} found six ultra-short ones attributable to planetary-mass objects between $10^{-6}$ and $10^{-4}\,M_{\odot}$. These would contribute $\Ocal( 1\% )$ of the dark matter.\\[2mm]
{\it Pulsar Timing{\;---\;}}Recently, NANOGrav has detected a stochastic signal in the time residuals from their $12.5\,$year pulsar-timing array data~\citep{Arzoumanian:2020vkk}. Several groups~\citep{Kohri:2020qqd, DeLuca:2020agl, Vaskonen:2020lbd, Domenech:2020ers, 2020arXiv201005071B} have attributed this to a stochastic background of gravitational waves from planetary-mass PBHs, this being consistent with the short timescale microlensing events found in OGLE data.\\[2mm]
{\it Quasar Microlensing{\;---\;}} The detection of $24$ microlensed quasars by~\cite{Mediavilla:2017bok} would allow up to $25\%$ of galactic halos to be PBHs in the mass range $0.05$ to $0.45\,M_\odot$. The microlensing could also be explained by intervening stars, but in several cases the stellar region of the lensing galaxy is not aligned with the quasar, which suggests a population of subsolar objects with $f_{\rm PBH} > 0.01$. A related claim was previously made by~\cite{Hawkins:2006xj}.\\[2mm]
{\it Cosmic Infrared/X-ray Backgrounds{\;---\;}}~\cite{2005Natur.438...45K} and~\cite{Kashlinsky:2016sdv} have suggested that the spatial coherence of the X-ray and infrared source-subtracted backgrounds could be explained by a significant density of PBHs larger than $\Ocal( 1 )\,M_{\odot}$, the Poisson fluctuations in their number density then producing halos earlier than usual. In such halos, a few stars form and emit infrared radiation, while the PBHs emit X-rays due to accretion.
\\[1mm]
{\it Ultra-Faint Dwarf Galaxies{\;---\;}}The non-detection of galaxies smaller than $10\text{--}20$~parsecs, despite their magnitude being above the detection limit, suggests compact halo objects in the solar-mass range. Moreover, rapid accretion in the densest PBH halos could explain the observed extreme Ultra-Faint Dwarf Galaxies mass-to-light ratios~\citep{Clesse:2017bsw}. Recent $N$-body simulations~\citep{Boldrini:2019isx} support this suggestion if PBHs in the mass range $25\text{--}100\,M_{\odot}$ provide at least $1\%$ of the dark matter.\\[1mm]

If confirmed, any of these claimed signatures would rule out the thermal WIMP model considered here from providing a significant fraction of the DM. However, our analysis has disregarded the possibility that WIMP halos could be dynamically disrupted if PBHs provide most of the DM, as discussed by~\citet{Hertzberg:2019exb}, and this might modify our conclusion.

\section{Discussions and Outlook}
\label{sec:Discusson-and-Outlook}

In this work, we have examined the bounds on the WIMP and PBH DM fractions from WIMP annihilations around PBHs with masses from $10^{-18}$ to $10^{15}\,M_{\odot}$. Our results are summarised in Fig.~\ref{fig:fPBH} for the case $f_{\rm PBH} \lesssim f_{\chi}$ and in Fig.~\ref{fig:fchi} for the case $f_{\chi} \lesssim f_{\rm PBH}$.

We have first studied the effects of DM annihilation when the dominant DM component is WIMPs from thermal freeze-out. For PBHs larger than a planetary mass, the expression for the extragalactic $\gamma$-ray flux in Eq.~\eqref{eq:flux} is independent of $M$, so the effect of the PBH mass function is unimportant and the maximally-allowed PBH DM fraction is $f^{}_{\rm PBH} \lesssim 2 \times 10^{-9}\,( m_{\chi}/ {\rm TeV} )^{1.1}$ for $M \lesssim 5 \times 10^{12}\,M_{\odot}\,( m_{\chi}/ {\rm TeV} )^{1.1}$. However, the limit on $f_{\rm PBH}$ is a decreasing function of mass for small $M$, so one could have a significant density of both WIMPs and PBHs for $M$ in the asteroid mass range.

We also studied the effects of DM annihilations when the dominant DM component is PBHs. This is particularly relevant for the merging intermediate-mass black holes recently discovered by the LIGO/Virgo collaboration~\citep{Abbott:2020tfl, Abbott:2020mjq}. In particular, the collaboration has reported a gravitational-wave signal consistent with a black hole binary with component masses of $85^{+21}_{-14}\,M_{\odot}$ and $66^{+17}_{-18}\,M_{\odot}$. It is hard to form black holes from stellar evolution in this range~\citep{Belczynski:2016jno, Spera:2017fyx}, so this could indicate that the components were of primordial origin. The LIGO/Virgo black holes may not provide all the DM but they must provide at least $1\%$ of it. However, Fig.~\ref{fig:fchi} shows that even this would rule out the standard WIMP DM scenario, so this may require a {\it third} DM component (the `something else' of our title). 

Alternatively, the PBHs could have an extended mass function, so that $f_{\rm PBH} = 1$ even if the LIGO/Virgo black holes have a much smaller density. In this case, the LIGO/Virgo discovery signals a paradigm shift from microscopic to macroscopic DM. If the PBH mass spectrum is dictated by the thermal history of the Universe, this could solve several other cosmic conundra~\citep{Carr:2019kxo}.

Our limits are weakened if part of the Galactic and extragalactic backgrounds is generated by some other source. For example, part of the extragalactic background can be attributed to TeV blazars~\citep{2009ApJ...707.1310A, 2010Sci...328...73N, Ghisellini:2017ico} and part of the Galactic background might come from to DM subhalos~\citep{Calore:2016ogv} or evaporating PBHs~\citep{Fermi-LAT:2018pfs}. Also superradiant spinning BHs could interact with accreting gas to generate another sort of BH $\gamma$-ray halo~\citep{2021arXiv210211280C}. Inverse Compton scattering in the accretion disk around a BH produces photons in the keV but not $\gamma$-ray range~\citep{Sunyaev:1979nz}.

Astrophysical processes could modify the WIMP halos around BHs within galaxies. In particular, tidal stripping could modify the halos in this case. When a star passes near a BH, it deposits energy into the halo, which could remove part of it~\citep{Green:2006hh}. This mechanism has been invoked for self-gravitating halos made of WIMPs~\citep{Schneider:2010jr} or axions~\citep{Tinyakov:2015cgg, 2020arXiv201105377K} and it has recently been applied to WIMP halos around BHs~\citep{Hertzberg:2019exb}. Of course, part of the halo could also be removed by the interaction amongst PBH-halo systems, particularly in high-density regions such as in the galactic centres or in PBH clusters. We leave this for future work.

For an individual BH, the maximum distance where $\gamma$-rays from DM annihilation can be detected is~\citep{Carr:2020erq}
\begin{equation}
	\label{eq:effectived}
	d_{\Lrm}
		 =
					\sqrt{\,
					\frac{ \Gamma\,N_{\gamma}( m_{\chi} ) }
					{ 2\,\Phi_{\rm Fermi}} }
		\approx
					1.4 {\rm\,kpc}
					\left(
						\frac{ M }{ M_{\odot} }
					\right)^{\!0.5}
					\left(
						\frac{ m_{\chi} }{ \rm TeV } 
					\right)^{0.82}
					,
\end{equation}
where we used Eqs.~(\ref{eq:Gamma},\.\ref{eq:upsilon}) at the last step. PBHs of sub-solar mass would mostly be visible within $2\,$kpc of the Solar System, where WIMP halos should not be subject to tidal stripping. However, the assessment of this effect is very dependent on $f_{\rm PBH}$, $M$, and the orbital radius. Future Monte Carlo numerical simulations could be used to estimate the BH population in the Galaxy as a function of orbital radius and in the Universe as a function of redshift.

Our analysis can be improved by dropping some of the assumptions made. 
	({\it i}$\mspace{2mu}$) We have assumed the WIMP cross-section does not change during 
		the evolution of the Universe but this is not true if a light mediator leads to 
		a Sommerfeld enhancement of the WIMP annihilation~\citep{ArkaniHamed:2008qn}.
	({\it ii}$\mspace{2mu}$) The cross-section has been fixed to the value required at 
		freeze-out with standard cosmology but its value might vary considerably in 
		non-standard cosmologies [e.g.\ with an early period of matter dominance or some other 
		exotic equation of state~\citep{Gelmini:2008sh}].
	({\it iii}$\mspace{2mu}$) We have assumed $s$-channel annihilation but
		the expected signal from annihilations must be reconsidered if the WIMP velocity distribution 
		plays a r{\^o}le in the computation of $\sv$ (e.g.\ if corrections of order $( v / c )^{2}$ 
		are to be taken into account). Thus our analysis does not cover other important particle DM candidates, 
		such as the sterile neutrino~\citep{Boyarsky:2018tvu} or the QCD axion~\citep{DiLuzio:2020wdo}. 
		If exotic particles do form gravitationally bound structures around black holes, they would provide 
		a powerful cosmological test due to their unique imprints.
		\vs{2mm}

{\bf Note added:} Just before submission of this revised version of our paper, a preprint by~\cite{Boudaud:2021irr} appeared with a similar analysis of the radial profile of the WIMP distribution to that presented below. This work was done independently but our Eq.~\eqref{eq:densityprofile} shows the same three power-law regimes as Boudaud {\it et al.} Although there was a mistake the Appendix of the earlier version of our paper, this does not affect our constraints on the mass of the PBHs or WIMPs since the initial profile is erased by annihilations.

\section*{Acknowledgements}
We thank Bradley Kavanagh and the referee for helpful comments. F.K.~acknowledges hospitality and support from the Delta Institute for Theoretical Physics. L.V.~acknowledges support from the NWO Physics Vrij Programme ``The Hidden Universe of Weakly Interacting Particles'' with project number 680.92.18.03 (NWO Vrije Programma), which is (partly) financed by the Dutch Research Council (NWO), as well as support from the European Union's Horizon 2020 research and innovation programme under the Marie Sk{\l}odowska-Curie grant agreement No.~754496 (H2020-MSCA-COFUND-2016 FELLINI).

\section*{Data Availability}
No new data were generated or analysed in support of this research.

\bibliographystyle{mnras}
\bibliography{PBHbib}

\begin{thebibliography}{}
\makeatletter
\relax
\def\mn@urlcharsother{\let\do\@makeother \do\$\do\&\do\#\do\^\do\_\do\%\do\~}
\def\mn@doi{\begingroup\mn@urlcharsother \@ifnextchar [ {\mn@doi@}
  {\mn@doi@[]}}
\def\mn@doi@[#1]#2{\def\@tempa{#1}\ifx\@tempa\@empty \href
  {http://dx.doi.org/#2} {doi:#2}\else \href {http://dx.doi.org/#2} {#1}\fi
  \endgroup}
\def\mn@eprint#1#2{\mn@eprint@#1:#2::\@nil}
\def\mn@eprint@arXiv#1{\href {http://arxiv.org/abs/#1} {{\tt arXiv:#1}}}
\def\mn@eprint@dblp#1{\href {http://dblp.uni-trier.de/rec/bibtex/#1.xml}
  {dblp:#1}}
\def\mn@eprint@#1:#2:#3:#4\@nil{\def\@tempa {#1}\def\@tempb {#2}\def\@tempc
  {#3}\ifx \@tempc \@empty \let \@tempc \@tempb \let \@tempb \@tempa \fi \ifx
  \@tempb \@empty \def\@tempb {arXiv}\fi \@ifundefined
  {mn@eprint@\@tempb}{\@tempb:\@tempc}{\expandafter \expandafter \csname
  mn@eprint@\@tempb\endcsname \expandafter{\@tempc}}}

\bibitem[\protect\citeauthoryear{Abbott \& Sikivie}{Abbott \&
  Sikivie}{1983}]{Abbott:1982af}
Abbott L.~F.,  Sikivie P.,  1983, \mn@doi [Phys. Lett.]
  {10.1016/0370-2693(83)90638-X}, B120, 133

\bibitem[\protect\citeauthoryear{Abbott et~al.}{Abbott
  et~al.}{2020a}]{PhysRevLett.125.101102}
Abbott R.,  et~al., 2020a, \mn@doi [Phys. Rev. Lett.]
  {10.1103/PhysRevLett.125.101102}, 125, 101102

\bibitem[\protect\citeauthoryear{Abbott et~al.}{Abbott
  et~al.}{2020b}]{Abbott:2020tfl}
Abbott R.,  et~al., 2020b, \mn@doi [Phys. Rev. Lett.]
  {10.1103/PhysRevLett.125.101102}, 125, 101102

\bibitem[\protect\citeauthoryear{Abbott et~al.}{Abbott
  et~al.}{2020c}]{Abbott:2020mjq}
Abbott R.,  et~al., 2020c, \mn@doi [Astrophys. J.] {10.3847/2041-8213/aba493},
  900, L13

\bibitem[\protect\citeauthoryear{{Abdo} et~al.,}{{Abdo}
  et~al.}{2009}]{2009ApJ...707.1310A}
{Abdo} A.~A.,  et~al., 2009, \mn@doi [\apj] {10.1088/0004-637X/707/2/1310},
  \href {https://ui.adsabs.harvard.edu/abs/2009ApJ...707.1310A} {707, 1310}

\bibitem[\protect\citeauthoryear{{Abdo} et~al.}{{Abdo}
  et~al.}{2010}]{2010PhRvL.104j1101A}
{Abdo} A.~A.,  et~al., 2010, \mn@doi [\prl] {10.1103/PhysRevLett.104.101101},
  \href {https://ui.adsabs.harvard.edu/abs/2010PhRvL.104j1101A} {104, 101101}

\bibitem[\protect\citeauthoryear{Acharya, Kane, Watson  \& Kumar}{Acharya
  et~al.}{2009}]{Acharya:2009zt}
Acharya B.~S.,  Kane G.,  Watson S.,   Kumar P.,  2009, \mn@doi [Phys. Rev. D]
  {10.1103/PhysRevD.80.083529}, 80, 083529

\bibitem[\protect\citeauthoryear{Acharya et~al.}{Acharya
  et~al.}{2018}]{CTAConsortium:2018tzg}
Acharya B.,  et~al., 2018, {Science with the Cherenkov Telescope Array}.
WSP (\mn@eprint {arXiv} {1709.07997}), \mn@doi{10.1142/10986}

\bibitem[\protect\citeauthoryear{Ackermann et~al.}{Ackermann
  et~al.}{2015}]{Ackermann:2015tah}
Ackermann M.,  et~al., 2015, \mn@doi [JCAP] {10.1088/1475-7516/2015/09/008},
  09, 008

\bibitem[\protect\citeauthoryear{Ackermann et~al.}{Ackermann
  et~al.}{2018}]{Fermi-LAT:2018pfs}
Ackermann M.,  et~al., 2018, \mn@doi [Astrophys. J.]
  {10.3847/1538-4357/aaac7b}, 857, 49

\bibitem[\protect\citeauthoryear{Adamek, Byrnes, Gosenca  \& Hotchkiss}{Adamek
  et~al.}{2019}]{Adamek:2019gns}
Adamek J.,  Byrnes C.~T.,  Gosenca M.,   Hotchkiss S.,  2019, \mn@doi [Phys.
  Rev.] {10.1103/PhysRevD.100.023506}, D100, 023506

\bibitem[\protect\citeauthoryear{Aghanim et~al.}{Aghanim
  et~al.}{2020}]{Aghanim:2018eyx}
Aghanim N.,  et~al., 2020, \mn@doi [Astron. Astrophys.]
  {10.1051/0004-6361/201833910}, 641, A6

\bibitem[\protect\citeauthoryear{Amoroso, Caron, Jueid, Ruiz~de Austri  \&
  Skands}{Amoroso et~al.}{2019}]{Amoroso:2018qga}
Amoroso S.,  Caron S.,  Jueid A.,  Ruiz~de Austri R.,   Skands P.,  2019,
  \mn@doi [JCAP] {10.1088/1475-7516/2019/05/007}, 05, 007

\bibitem[\protect\citeauthoryear{Ando \& Ishiwata}{Ando \&
  Ishiwata}{2015}]{Ando:2015qda}
Ando S.,  Ishiwata K.,  2015, \mn@doi [JCAP] {10.1088/1475-7516/2015/05/024},
  05, 024

\bibitem[\protect\citeauthoryear{Arkani-Hamed, Finkbeiner, Slatyer  \&
  Weiner}{Arkani-Hamed et~al.}{2009}]{ArkaniHamed:2008qn}
Arkani-Hamed N.,  Finkbeiner D.~P.,  Slatyer T.~R.,   Weiner N.,  2009, \mn@doi
  [Phys. Rev. D] {10.1103/PhysRevD.79.015014}, 79, 015014

\bibitem[\protect\citeauthoryear{Arzoumanian et~al.}{Arzoumanian
  et~al.}{2020}]{Arzoumanian:2020vkk}
Arzoumanian Z.,  et~al., 2020, \mn@doi [Astrophys. J. Lett.]
  {10.3847/2041-8213/abd401}, 905, L34

\bibitem[\protect\citeauthoryear{Bai et~al.}{Bai et~al.}{2019}]{Bai:2019khm}
Bai X.,  et~al., 2019, arXiv e-prints, \href
  {https://ui.adsabs.harvard.edu/abs/2019arXiv190502773B} {p. arXiv:1905.02773}

\bibitem[\protect\citeauthoryear{Baum, Visinelli, Freese  \& Stengel}{Baum
  et~al.}{2017}]{Baum:2016oow}
Baum S.,  Visinelli L.,  Freese K.,   Stengel P.,  2017, \mn@doi [Phys.\ Rev.\
  D] {10.1103/PhysRevD.95.043007}, 95, 043007

\bibitem[\protect\citeauthoryear{B{\'e}langer, Boudjema, Goudelis, Pukhov  \&
  Zaldivar}{B{\'e}langer et~al.}{2018}]{Belanger:2018ccd}
B{\'e}langer G.,  Boudjema F.,  Goudelis A.,  Pukhov A.,   Zaldivar B.,  2018,
  \mn@doi [Comput. Phys. Commun.] {10.1016/j.cpc.2018.04.027}, 231, 173

\bibitem[\protect\citeauthoryear{Belczynski et~al.}{Belczynski
  et~al.}{2016}]{Belczynski:2016jno}
Belczynski K.,  et~al., 2016, \mn@doi [Astron. Astrophys.]
  {10.1051/0004-6361/201628980}, 594, A97

\bibitem[\protect\citeauthoryear{Berezinsky, Gurevich  \& Zybin}{Berezinsky
  et~al.}{1992}]{Berezinsky:1992mx}
Berezinsky V.,  Gurevich A.,   Zybin K.,  1992, \mn@doi [Phys. Lett. B]
  {10.1016/0370-2693(92)90686-X}, 294, 221

\bibitem[\protect\citeauthoryear{Bernstein, Brown  \& Feinberg}{Bernstein
  et~al.}{1985}]{Bernstein:1985th}
Bernstein J.,  Brown L.~S.,   Feinberg G.,  1985, \mn@doi [Phys. Rev. D]
  {10.1103/PhysRevD.32.3261}, 32, 3261

\bibitem[\protect\citeauthoryear{Bertone, Coogan, Gaggero, Kavanagh  \&
  Weniger}{Bertone et~al.}{2019}]{Bertone:2019vsk}
Bertone G.,  Coogan A.~M.,  Gaggero D.,  Kavanagh B.~J.,   Weniger C.,  2019,
  \mn@doi [Phys. Rev. D] {10.1103/PhysRevD.100.123013}, 100, 123013

\bibitem[\protect\citeauthoryear{{Bhattacharya}, {Mohanty}  \&
  {Parashari}}{{Bhattacharya} et~al.}{2020}]{2020arXiv201005071B}
{Bhattacharya} S.,  {Mohanty} S.,   {Parashari} P.,  2020, arXiv e-prints,
  \href {https://ui.adsabs.harvard.edu/abs/2020arXiv201005071B} {p.
  arXiv:2010.05071}

\bibitem[\protect\citeauthoryear{Boldrini, Miki, Wagner, Mohayaee, Silk  \&
  Arbey}{Boldrini et~al.}{2020}]{Boldrini:2019isx}
Boldrini P.,  Miki Y.,  Wagner A.~Y.,  Mohayaee R.,  Silk J.,   Arbey A.,
  2020, \mn@doi [Mon. Not. Roy. Astron. Soc.] {10.1093/mnras/staa150}, 492,
  5218

\bibitem[\protect\citeauthoryear{Boucenna, Kuhnel, Ohlsson  \&
  Visinelli}{Boucenna et~al.}{2018}]{Boucenna:2017ghj}
Boucenna S.~M.,  Kuhnel F.,  Ohlsson T.,   Visinelli L.,  2018, \mn@doi [JCAP]
  {10.1088/1475-7516/2018/07/003}, 1807, 003

\bibitem[\protect\citeauthoryear{{Boudaud}, {Lacroix}, {Stref}, {Lavalle}  \&
  {Salati}}{{Boudaud} et~al.}{2021}]{Boudaud:2021irr}
{Boudaud} M.,  {Lacroix} T.,  {Stref} M.,  {Lavalle} J.,   {Salati} P.,  2021,
  arXiv e-prints, \href {https://ui.adsabs.harvard.edu/abs/2021arXiv210607480B}
  {p. arXiv:2106.07480}

\bibitem[\protect\citeauthoryear{Boyarsky, Drewes, Lasserre, Mertens  \&
  Ruchayskiy}{Boyarsky et~al.}{2019}]{Boyarsky:2018tvu}
Boyarsky A.,  Drewes M.,  Lasserre T.,  Mertens S.,   Ruchayskiy O.,  2019,
  \mn@doi [Prog. Part. Nucl. Phys.] {10.1016/j.ppnp.2018.07.004}, 104, 1

\bibitem[\protect\citeauthoryear{Bringmann \& Hofmann}{Bringmann \&
  Hofmann}{2007}]{Bringmann:2006mu}
Bringmann T.,  Hofmann S.,  2007, \mn@doi [JCAP]
  {10.1088/1475-7516/2007/04/016}, 04, 016

\bibitem[\protect\citeauthoryear{Bringmann, Scott  \& Akrami}{Bringmann
  et~al.}{2012}]{Bringmann:2011ut}
Bringmann T.,  Scott P.,   Akrami Y.,  2012, \mn@doi [Phys. Rev. D]
  {10.1103/PhysRevD.85.125027}, 85, 125027

\bibitem[\protect\citeauthoryear{Bringmann, Edsj{\"o}, Gondolo, Ullio  \&
  Bergstr{\"o}m}{Bringmann et~al.}{2018}]{Bringmann:2018lay}
Bringmann T.,  Edsj{\"o} J.,  Gondolo P.,  Ullio P.,   Bergstr{\"o}m L.,  2018,
  \mn@doi [JCAP] {10.1088/1475-7516/2018/07/033}, 07, 033

\bibitem[\protect\citeauthoryear{{Cai}, {Yang}  \& {Zhou}}{{Cai}
  et~al.}{2020}]{Cai:2020fnq}
{Cai} R.-G.,  {Yang} X.-Y.,   {Zhou} Y.-F.,  2020, arXiv e-prints, \href
  {https://ui.adsabs.harvard.edu/abs/2020arXiv200711804C} {p. arXiv:2007.11804}

\bibitem[\protect\citeauthoryear{Calore, De~Romeri, Di~Mauro, Donato  \&
  Marinacci}{Calore et~al.}{2017}]{Calore:2016ogv}
Calore F.,  De~Romeri V.,  Di~Mauro M.,  Donato F.,   Marinacci F.,  2017,
  \mn@doi [Phys. Rev. D] {10.1103/PhysRevD.96.063009}, 96, 063009

\bibitem[\protect\citeauthoryear{{Caputo}, {Witte}, {Blas}  \& {Pani}}{{Caputo}
  et~al.}{2021}]{2021arXiv210211280C}
{Caputo} A.,  {Witte} S.~J.,  {Blas} D.,   {Pani} P.,  2021, arXiv e-prints,
  \href {https://ui.adsabs.harvard.edu/abs/2021arXiv210211280C} {p.
  arXiv:2102.11280}

\bibitem[\protect\citeauthoryear{Carr \& Hawking}{Carr \&
  Hawking}{1974}]{Carr:1974nx}
Carr B.~J.,  Hawking S.,  1974, Mon. Not. Roy. Astron. Soc., 168, 399

\bibitem[\protect\citeauthoryear{{Carr} \& {K{\"u}hnel}}{{Carr} \&
  {K{\"u}hnel}}{2020}]{Carr:2020xqk}
{Carr} B.,  {K{\"u}hnel} F.,  2020, \mn@doi [Annual Review of Nuclear and
  Particle Science] {10.1146/annurev-nucl-050520-125911}, \href
  {https://ui.adsabs.harvard.edu/abs/2020ARNPS..7050520C} {70, 355 }

\bibitem[\protect\citeauthoryear{Carr \& Sakellariadou}{Carr \&
  Sakellariadou}{1999}]{Carr:1997cn}
Carr B.~J.,  Sakellariadou M.,  1999, \mn@doi [Astrophys. J.] {10.1086/307071},
  516, 195

\bibitem[\protect\citeauthoryear{Carr, Kohri, Sendouda  \& Yokoyama}{Carr
  et~al.}{2010}]{Carr:2009jm}
Carr B.~J.,  Kohri K.,  Sendouda Y.,   Yokoyama J.,  2010, \mn@doi [Phys. Rev.]
  {10.1103/PhysRevD.81.104019}, D81, 104019

\bibitem[\protect\citeauthoryear{Carr, Kohri, Sendouda  \& Yokoyama}{Carr
  et~al.}{2016a}]{Carr:2016hva}
Carr B.,  Kohri K.,  Sendouda Y.,   Yokoyama J.,  2016a, \mn@doi [Phys. Rev. D]
  {10.1103/PhysRevD.94.044029}, 94, 044029

\bibitem[\protect\citeauthoryear{Carr, Kuhnel  \& Sandstad}{Carr
  et~al.}{2016b}]{Carr:2016drx}
Carr B.,  Kuhnel F.,   Sandstad M.,  2016b, \mn@doi [Phys.\ Rev.\ D]
  {10.1103/PhysRevD.94.083504}, 94, 083504

\bibitem[\protect\citeauthoryear{Carr, Raidal, Tenkanen, Vaskonen  \&
  Veerm\"ae}{Carr et~al.}{2017}]{Carr:2017jsz}
Carr B.,  Raidal M.,  Tenkanen T.,  Vaskonen V.,   Veerm\"ae H.,  2017, \mn@doi
  [Phys. Rev. D] {10.1103/PhysRevD.96.023514}, 96, 023514

\bibitem[\protect\citeauthoryear{Carr, Kuhnel  \& Visinelli}{Carr
  et~al.}{2020a}]{Carr:2020erq}
Carr B.,  Kuhnel F.,   Visinelli L.,  2020a, \mn@doi [Mon.\ Not.\ Roy.\
  Astron.\ Soc.] {https://doi.org/10.1093/mnras/staa3651}

\bibitem[\protect\citeauthoryear{Carr, Kohri, Sendouda  \& Yokoyama}{Carr
  et~al.}{2020b}]{Carr:2020gox}
Carr B.,  Kohri K.,  Sendouda Y.,   Yokoyama J.,  2020b, arXiv e-prints, \href
  {https://ui.adsabs.harvard.edu/abs/2020arXiv200212778C} {p. arXiv:2002.12778}

\bibitem[\protect\citeauthoryear{Carr, Clesse, Garcia-Bellido  \& Kuhnel}{Carr
  et~al.}{2021}]{Carr:2019kxo}
Carr B.,  Clesse S.,  Garcia-Bellido J.,   Kuhnel F.,  2021, \mn@doi [Phys.
  Dark Univ.] {10.1016/j.dark.2020.100755}, 31, 100755

\bibitem[\protect\citeauthoryear{{Chapline}}{{Chapline}}{1975}]{1975Natur.253..251C}
{Chapline} G.,  1975, \mn@doi [Nature (London)] {10.1038/253251a0}, \href
  {http://ads.nao.ac.jp/abs/1975Natur.253..251C} {253, 251}

\bibitem[\protect\citeauthoryear{Cirelli, Panci  \& Serpico}{Cirelli
  et~al.}{2010}]{Cirelli:2009dv}
Cirelli M.,  Panci P.,   Serpico P.~D.,  2010, \mn@doi [Nucl. Phys.]
  {10.1016/j.nuclphysb.2010.07.010}, B840, 284

\bibitem[\protect\citeauthoryear{Cirelli et~al.,}{Cirelli
  et~al.}{2011}]{Cirelli:2010xx}
Cirelli M.,  et~al., 2011, \mn@doi [JCAP] {10.1088/1475-7516/2012/10/E01,
  10.1088/1475-7516/2011/03/051}, 1103, 051

\bibitem[\protect\citeauthoryear{Clesse \& Garc\'ia-Bellido}{Clesse \&
  Garc\'ia-Bellido}{2018}]{Clesse:2017bsw}
Clesse S.,  Garc\'ia-Bellido J.,  2018, \mn@doi [Phys. Dark Univ.]
  {10.1016/j.dark.2018.08.004}, 22, 137

\bibitem[\protect\citeauthoryear{{Coogan}, {Morrison}  \& {Profumo}}{{Coogan}
  et~al.}{2020}]{Coogan:2020tuf}
{Coogan} A.,  {Morrison} L.,   {Profumo} S.,  2020, arXiv e-prints, \href
  {https://ui.adsabs.harvard.edu/abs/2020arXiv201004797C} {p. arXiv:2010.04797}

\bibitem[\protect\citeauthoryear{De~Luca, Franciolini  \& Riotto}{De~Luca
  et~al.}{2021}]{DeLuca:2020agl}
De~Luca V.,  Franciolini G.,   Riotto A.,  2021, \mn@doi [Phys. Rev. Lett.]
  {10.1103/PhysRevLett.126.041303}, 126, 041303

\bibitem[\protect\citeauthoryear{Di~Luzio, Giannotti, Nardi  \&
  Visinelli}{Di~Luzio et~al.}{2020}]{DiLuzio:2020wdo}
Di~Luzio L.,  Giannotti M.,  Nardi E.,   Visinelli L.,  2020, \mn@doi [Phys.
  Rept.] {10.1016/j.physrep.2020.06.002}, 870, 1

\bibitem[\protect\citeauthoryear{Di~Mauro \& Donato}{Di~Mauro \&
  Donato}{2015}]{DiMauro:2015tfa}
Di~Mauro M.,  Donato F.,  2015, \mn@doi [Phys. Rev. D]
  {10.1103/PhysRevD.91.123001}, 91, 123001

\bibitem[\protect\citeauthoryear{Di~Valentino, Melchiorri  \&
  Silk}{Di~Valentino et~al.}{2019a}]{DiValentino:2019qzk}
Di~Valentino E.,  Melchiorri A.,   Silk J.,  2019a, \mn@doi [Nat. Astron.]
  {10.1038/s41550-019-0906-9}, 4, 196

\bibitem[\protect\citeauthoryear{Di~Valentino, Ferreira, Visinelli  \&
  Danielsson}{Di~Valentino et~al.}{2019b}]{DiValentino:2019exe}
Di~Valentino E.,  Ferreira R.~Z.,  Visinelli L.,   Danielsson U.,  2019b,
  \mn@doi [Phys. Dark Univ.] {10.1016/j.dark.2019.100385}, 26, 100385

\bibitem[\protect\citeauthoryear{Dine \& Fischler}{Dine \&
  Fischler}{1983}]{Dine:1982ah}
Dine M.,  Fischler W.,  1983, \mn@doi [Phys. Lett.]
  {10.1016/0370-2693(83)90639-1}, B120, 137

\bibitem[\protect\citeauthoryear{Dodelson \& Widrow}{Dodelson \&
  Widrow}{1994}]{Dodelson:1993je}
Dodelson S.,  Widrow L.~M.,  1994, \mn@doi [Phys. Rev. Lett.]
  {10.1103/PhysRevLett.72.17}, 72, 17

\bibitem[\protect\citeauthoryear{{Dom{\`e}nech} \& {Pi}}{{Dom{\`e}nech} \&
  {Pi}}{2020}]{Domenech:2020ers}
{Dom{\`e}nech} G.,  {Pi} S.,  2020, arXiv e-prints, \href
  {https://ui.adsabs.harvard.edu/abs/2020arXiv201003976D} {p. arXiv:2010.03976}

\bibitem[\protect\citeauthoryear{Duda, Gelmini  \& Gondolo}{Duda
  et~al.}{2002}]{Duda:2001ae}
Duda G.,  Gelmini G.,   Gondolo P.,  2002, \mn@doi [Phys. Lett. B]
  {10.1016/S0370-2693(02)01266-2}, 529, 187

\bibitem[\protect\citeauthoryear{Eroshenko}{Eroshenko}{2016}]{Eroshenko:2016yve}
Eroshenko {\relax Yu}.~N.,  2016, \mn@doi [Astron. Lett.]
  {10.1134/S1063773716060013}, 42, 347

\bibitem[\protect\citeauthoryear{Eroshenko}{Eroshenko}{2020}]{Eroshenko:2019pxt}
Eroshenko Y.,  2020, \mn@doi [Int. J. Mod. Phys. A]
  {10.1142/S0217751X20400461}, 35, 2040046

\bibitem[\protect\citeauthoryear{{Franciolini} et~al.,}{{Franciolini}
  et~al.}{2021}]{Franciolini:2021tla}
{Franciolini} G.,  et~al., 2021, arXiv e-prints, \href
  {https://ui.adsabs.harvard.edu/abs/2021arXiv210503349F} {p. arXiv:2105.03349}

\bibitem[\protect\citeauthoryear{Gelmini \& Gondolo}{Gelmini \&
  Gondolo}{2006}]{Gelmini:2006pw}
Gelmini G.~B.,  Gondolo P.,  2006, \mn@doi [Phys. Rev. D]
  {10.1103/PhysRevD.74.023510}, 74, 023510

\bibitem[\protect\citeauthoryear{Gelmini \& Gondolo}{Gelmini \&
  Gondolo}{2008}]{Gelmini:2008sh}
Gelmini G.~B.,  Gondolo P.,  2008, \mn@doi [JCAP]
  {10.1088/1475-7516/2008/10/002}, 10, 002

\bibitem[\protect\citeauthoryear{Ghisellini, Righi, Costamante  \&
  Tavecchio}{Ghisellini et~al.}{2017}]{Ghisellini:2017ico}
Ghisellini G.,  Righi C.,  Costamante L.,   Tavecchio F.,  2017, \mn@doi [Mon.
  Not. Roy. Astron. Soc.] {10.1093/mnras/stx806}, 469, 255

\bibitem[\protect\citeauthoryear{Gondolo \& Gelmini}{Gondolo \&
  Gelmini}{1991}]{Gondolo:1990dk}
Gondolo P.,  Gelmini G.,  1991, \mn@doi [Nucl. Phys. B]
  {10.1016/0550-3213(91)90438-4}, 360, 145

\bibitem[\protect\citeauthoryear{Gondolo \& Silk}{Gondolo \&
  Silk}{1999}]{Gondolo:1999ef}
Gondolo P.,  Silk J.,  1999, \mn@doi [Phys. Rev. Lett.]
  {10.1103/PhysRevLett.83.1719}, 83, 1719

\bibitem[\protect\citeauthoryear{Green \& Goodwin}{Green \&
  Goodwin}{2007}]{Green:2006hh}
Green A.~M.,  Goodwin S.~P.,  2007, \mn@doi [Mon. Not. Roy. Astron. Soc.]
  {10.1111/j.1365-2966.2007.11397.x}, 375, 1111

\bibitem[\protect\citeauthoryear{Griest \& Seckel}{Griest \&
  Seckel}{1991}]{Griest:1990kh}
Griest K.,  Seckel D.,  1991, \mn@doi [Phys. Rev. D]
  {10.1103/PhysRevD.43.3191}, 43, 3191

\bibitem[\protect\citeauthoryear{Handley}{Handley}{2021}]{Handley:2019tkm}
Handley W.,  2021, \mn@doi [Phys. Rev. D] {10.1103/PhysRevD.103.L041301}, 103,
  L041301

\bibitem[\protect\citeauthoryear{Hawkins}{Hawkins}{2006}]{Hawkins:2006xj}
Hawkins M. R.~S.,  2006, \mn@doi [Astron. Astrophys.]
  {10.1051/0004-6361:20066283}

\bibitem[\protect\citeauthoryear{Hertzberg, Schiappacasse  \&
  Yanagida}{Hertzberg et~al.}{2020}]{Hertzberg:2019exb}
Hertzberg M.~P.,  Schiappacasse E.~D.,   Yanagida T.~T.,  2020, \mn@doi [Phys.
  Lett. B] {10.1016/j.physletb.2020.135566}, 807, 135566

\bibitem[\protect\citeauthoryear{Hooper \& Goodenough}{Hooper \&
  Goodenough}{2011}]{Hooper:2010mq}
Hooper D.,  Goodenough L.,  2011, \mn@doi [Phys. Lett. B]
  {10.1016/j.physletb.2011.02.029}, 697, 412

\bibitem[\protect\citeauthoryear{Hu, Barkana  \& Gruzinov}{Hu
  et~al.}{2000}]{Hu:2000ke}
Hu W.,  Barkana R.,   Gruzinov A.,  2000, \mn@doi [Phys. Rev. Lett.]
  {10.1103/PhysRevLett.85.1158}, 85, 1158

\bibitem[\protect\citeauthoryear{Hut}{Hut}{1977}]{Hut:1977zn}
Hut P.,  1977, \mn@doi [Phys. Lett. B] {10.1016/0370-2693(77)90139-3}, 69, 85

\bibitem[\protect\citeauthoryear{Josan \& Green}{Josan \&
  Green}{2010}]{Josan:2010vn}
Josan A.~S.,  Green A.~M.,  2010, \mn@doi [Phys. Rev.]
  {10.1103/PhysRevD.82.083527}, D82, 083527

\bibitem[\protect\citeauthoryear{Kadota \& Silk}{Kadota \&
  Silk}{2021}]{Kadota:2020ahr}
Kadota K.,  Silk J.,  2021, \mn@doi [Phys. Rev. D]
  {10.1103/PhysRevD.103.043530}, 103, 043530

\bibitem[\protect\citeauthoryear{Kashlinsky}{Kashlinsky}{2016}]{Kashlinsky:2016sdv}
Kashlinsky A.,  2016, \mn@doi [Astrophys.~J.] {10.3847/2041-8205/823/2/L25},
  823, L25

\bibitem[\protect\citeauthoryear{{Kashlinsky}, {Arendt}, {Mather}  \&
  {Moseley}}{{Kashlinsky} et~al.}{2005}]{2005Natur.438...45K}
{Kashlinsky} A.,  {Arendt} R.~G.,  {Mather} J.,   {Moseley} S.~H.,  2005,
  \mn@doi [\nat] {10.1038/nature04143}, \href
  {https://ui.adsabs.harvard.edu/abs/2005Natur.438...45K} {438, 45}

\bibitem[\protect\citeauthoryear{{Kavanagh}, {Edwards}, {Visinelli}  \&
  {Weniger}}{{Kavanagh} et~al.}{2020}]{2020arXiv201105377K}
{Kavanagh} B.~J.,  {Edwards} T. D.~P.,  {Visinelli} L.,   {Weniger} C.,  2020,
  arXiv e-prints, \href {https://ui.adsabs.harvard.edu/abs/2020arXiv201105377K}
  {p. arXiv:2011.05377}

\bibitem[\protect\citeauthoryear{Kohri \& Terada}{Kohri \&
  Terada}{2021}]{Kohri:2020qqd}
Kohri K.,  Terada T.,  2021, \mn@doi [Phys. Lett. B]
  {10.1016/j.physletb.2020.136040}, 813, 136040

\bibitem[\protect\citeauthoryear{K{\"u}hnel \& Freese}{K{\"u}hnel \&
  Freese}{2017}]{Kuhnel:2017pwq}
K{\"u}hnel F.,  Freese K.,  2017, \mn@doi [Phys. Rev. D]
  {10.1103/PhysRevD.95.083508}, 95, 083508

\bibitem[\protect\citeauthoryear{K{\"u}hnel, Rampf  \& Sandstad}{K{\"u}hnel
  et~al.}{2016}]{Kuhnel:2015vtw}
K{\"u}hnel F.,  Rampf C.,   Sandstad M.,  2016, \mn@doi [Eur. Phys. J. C]
  {10.1140/epjc/s10052-016-3945-8}, 76, 93

\bibitem[\protect\citeauthoryear{Lacki \& Beacom}{Lacki \&
  Beacom}{2010}]{Lacki:2010zf}
Lacki B.~C.,  Beacom J.~F.,  2010, \mn@doi [Astrophys. J.]
  {10.1088/2041-8205/720/1/L67}, 720, L67

\bibitem[\protect\citeauthoryear{Lee \& Weinberg}{Lee \&
  Weinberg}{1977}]{Lee:1977ua}
Lee B.~W.,  Weinberg S.,  1977, \mn@doi [Phys. Rev. Lett.]
  {10.1103/PhysRevLett.39.165}, 39, 165

\bibitem[\protect\citeauthoryear{Mack, Ostriker  \& Ricotti}{Mack
  et~al.}{2007}]{Mack:2006gz}
Mack K.~J.,  Ostriker J.~P.,   Ricotti M.,  2007, \mn@doi [Astrophys. J.]
  {10.1086/518998}, 665, 1277

\bibitem[\protect\citeauthoryear{Mediavilla, Jim\'enez-Vicente, Mu\~noz,
  Vives-Arias  \& Calder\'on-Infante}{Mediavilla
  et~al.}{2017}]{Mediavilla:2017bok}
Mediavilla E.,  Jim\'enez-Vicente J.,  Mu\~noz J.~A.,  Vives-Arias H.,
  Calder\'on-Infante J.,  2017, \mn@doi [Astrophys. J.]
  {10.3847/2041-8213/aa5dab}, 836, L18

\bibitem[\protect\citeauthoryear{{Mr{\'o}z} et~al.}{{Mr{\'o}z}
  et~al.}{2017}]{2017Natur.548..183M}
{Mr{\'o}z} P.,  et~al., 2017, \mn@doi [\nat] {10.1038/nature23276}, \href
  {https://ui.adsabs.harvard.edu/abs/2017Natur.548..183M} {548, 183}

\bibitem[\protect\citeauthoryear{{Neronov} \& {Vovk}}{{Neronov} \&
  {Vovk}}{2010}]{2010Sci...328...73N}
{Neronov} A.,  {Vovk} I.,  2010, \mn@doi [Science] {10.1126/science.1184192},
  \href {https://ui.adsabs.harvard.edu/abs/2010Sci...328...73N} {328, 73}

\bibitem[\protect\citeauthoryear{Niikura, Takada, Yokoyama, Sumi  \&
  Masaki}{Niikura et~al.}{2019}]{Niikura:2019kqi}
Niikura H.,  Takada M.,  Yokoyama S.,  Sumi T.,   Masaki S.,  2019, \mn@doi
  [Phys. Rev.] {10.1103/PhysRevD.99.083503}, D99, 083503

\bibitem[\protect\citeauthoryear{{Peebles}}{{Peebles}}{1972}]{1972GReGr...3...63P}
{Peebles} P.~J.~E.,  1972, \mn@doi [Gen.~Relativ.~Gravit.]
  {10.1007/BF00755923}, \href
  {https://ui.adsabs.harvard.edu/abs/1972GReGr...3...63P} {3, 63}

\bibitem[\protect\citeauthoryear{{Phukon} et~al.,}{{Phukon}
  et~al.}{2021}]{Phukon:2021cus}
{Phukon} K.~S.,  et~al., 2021, arXiv e-prints, \href
  {https://ui.adsabs.harvard.edu/abs/2021arXiv210511449P} {p. arXiv:2105.11449}

\bibitem[\protect\citeauthoryear{Poulin, Smith, Karwal  \& Kamionkowski}{Poulin
  et~al.}{2019}]{Poulin:2018cxd}
Poulin V.,  Smith T.,  Karwal T.,   Kamionkowski M.,  2019, \mn@doi [Phys. Rev.
  Lett.] {10.1103/PhysRevLett.122.221301}, 122, 221301

\bibitem[\protect\citeauthoryear{Preskill, Wise  \& Wilczek}{Preskill
  et~al.}{1983}]{Preskill:1982cy}
Preskill J.,  Wise M.~B.,   Wilczek F.,  1983, \mn@doi [Phys. Lett.]
  {10.1016/0370-2693(83)90637-8}, B120, 127

\bibitem[\protect\citeauthoryear{{Quinlan}, {Hernquist}  \&
  {Sigurdsson}}{{Quinlan} et~al.}{1995}]{1995ApJ...440..554Q}
{Quinlan} G.~D.,  {Hernquist} L.,   {Sigurdsson} S.,  1995, \mn@doi
  [Astrophys.~J.] {10.1086/175295}, \href
  {https://ui.adsabs.harvard.edu/abs/1995ApJ...440..554Q} {440, 554}

\bibitem[\protect\citeauthoryear{Ricotti}{Ricotti}{2007}]{Ricotti:2007jk}
Ricotti M.,  2007, \mn@doi [Astrophys. J.] {10.1086/516562}, 662, 53

\bibitem[\protect\citeauthoryear{Ricotti \& Gould}{Ricotti \&
  Gould}{2009}]{Ricotti:2009bs}
Ricotti M.,  Gould A.,  2009, \mn@doi [Astrophys. J.]
  {10.1088/0004-637X/707/2/979}, 707, 979

\bibitem[\protect\citeauthoryear{Ricotti, Ostriker  \& Mack}{Ricotti
  et~al.}{2008}]{Ricotti:2007au}
Ricotti M.,  Ostriker J.~P.,   Mack K.~J.,  2008, \mn@doi [Astrophys. J.]
  {10.1086/587831}, 680, 829

\bibitem[\protect\citeauthoryear{Saito \& Shirai}{Saito \&
  Shirai}{2011}]{Saito:2010ts}
Saito R.,  Shirai S.,  2011, \mn@doi [Phys. Lett. B]
  {10.1016/j.physletb.2011.01.038}, 697, 95

\bibitem[\protect\citeauthoryear{Sato \& Kobayashi}{Sato \&
  Kobayashi}{1977}]{Sato:1977ye}
Sato K.,  Kobayashi M.,  1977, \mn@doi [Prog. Theor. Phys.]
  {10.1143/PTP.58.1775}, 58, 1775

\bibitem[\protect\citeauthoryear{Schive, Chiueh  \& Broadhurst}{Schive
  et~al.}{2014}]{Schive:2014dra}
Schive H.-Y.,  Chiueh T.,   Broadhurst T.,  2014, \mn@doi [Nature Phys.]
  {10.1038/nphys2996}, 10, 496

\bibitem[\protect\citeauthoryear{Schneider, Krauss  \& Moore}{Schneider
  et~al.}{2010}]{Schneider:2010jr}
Schneider A.,  Krauss L.,   Moore B.,  2010, \mn@doi [Phys. Rev. D]
  {10.1103/PhysRevD.82.063525}, 82, 063525

\bibitem[\protect\citeauthoryear{Scott \& Sivertsson}{Scott \&
  Sivertsson}{2009}]{Scott:2009tu}
Scott P.,  Sivertsson S.,  2009, \mn@doi [Phys. Rev. Lett.]
  {10.1103/PhysRevLett.105.119902, 10.1103/PhysRevLett.103.211301}, 103, 211301

\bibitem[\protect\citeauthoryear{Shemmer, Netzer, Maiolino, Oliva, Croom,
  Corbett  \& di Fabrizio}{Shemmer et~al.}{2004}]{Shemmer:2004ph}
Shemmer O.,  Netzer H.,  Maiolino R.,  Oliva E.,  Croom S.,  Corbett E.,   di
  Fabrizio L.,  2004, \mn@doi [Astrophys. J.] {10.1086/423607}, 614, 547

\bibitem[\protect\citeauthoryear{Shi \& Fuller}{Shi \&
  Fuller}{1999}]{Shi:1998km}
Shi X.-D.,  Fuller G.~M.,  1999, \mn@doi [Phys. Rev. Lett.]
  {10.1103/PhysRevLett.82.2832}, 82, 2832

\bibitem[\protect\citeauthoryear{Slatyer, Padmanabhan  \& Finkbeiner}{Slatyer
  et~al.}{2009}]{Slatyer:2009yq}
Slatyer T.~R.,  Padmanabhan N.,   Finkbeiner D.~P.,  2009, \mn@doi [Phys. Rev.]
  {10.1103/PhysRevD.80.043526}, D80, 043526

\bibitem[\protect\citeauthoryear{Spera \& Mapelli}{Spera \&
  Mapelli}{2017}]{Spera:2017fyx}
Spera M.,  Mapelli M.,  2017, \mn@doi [Mon. Not. Roy. Astron. Soc.]
  {10.1093/mnras/stx1576}, 470, 4739

\bibitem[\protect\citeauthoryear{Steigman}{Steigman}{1979}]{Steigman:1979kw}
Steigman G.,  1979, \mn@doi [Ann. Rev. Nucl. Part. Sci.]
  {10.1146/annurev.ns.29.120179.001525}, 29, 313

\bibitem[\protect\citeauthoryear{Steigman, Dasgupta  \& Beacom}{Steigman
  et~al.}{2012}]{Steigman:2012nb}
Steigman G.,  Dasgupta B.,   Beacom J.~F.,  2012, \mn@doi [Phys. Rev. D]
  {10.1103/PhysRevD.86.023506}, 86, 023506

\bibitem[\protect\citeauthoryear{Sunyaev \& Truemper}{Sunyaev \&
  Truemper}{1979}]{Sunyaev:1979nz}
Sunyaev R.~A.,  Truemper J.,  1979, \mn@doi [Nature] {10.1038/279506a0}, 279,
  506

\bibitem[\protect\citeauthoryear{Tinyakov, Tkachev  \& Zioutas}{Tinyakov
  et~al.}{2016}]{Tinyakov:2015cgg}
Tinyakov P.,  Tkachev I.,   Zioutas K.,  2016, \mn@doi [JCAP]
  {10.1088/1475-7516/2016/01/035}, 01, 035

\bibitem[\protect\citeauthoryear{Ullio, Zhao  \& Kamionkowski}{Ullio
  et~al.}{2001}]{Ullio:2001fb}
Ullio P.,  Zhao H.,   Kamionkowski M.,  2001, \mn@doi [Phys. Rev. D]
  {10.1103/PhysRevD.64.043504}, 64, 043504

\bibitem[\protect\citeauthoryear{Ullio, Bergstrom, Edsjo  \& Lacey}{Ullio
  et~al.}{2002}]{Ullio:2002pj}
Ullio P.,  Bergstrom L.,  Edsjo J.,   Lacey C.~G.,  2002, \mn@doi [Phys. Rev.]
  {10.1103/PhysRevD.66.123502}, D66, 123502

\bibitem[\protect\citeauthoryear{{Vagnozzi}, {Di Valentino}, {Gariazzo},
  {Melchiorri}, {Mena}  \& {Silk}}{{Vagnozzi} et~al.}{2020}]{Vagnozzi:2020zrh}
{Vagnozzi} S.,  {Di Valentino} E.,  {Gariazzo} S.,  {Melchiorri} A.,  {Mena}
  O.,   {Silk} J.,  2020, arXiv e-prints, \href
  {https://ui.adsabs.harvard.edu/abs/2020arXiv201002230V} {p. arXiv:2010.02230}

\bibitem[\protect\citeauthoryear{Vaskonen \& Veerm\"ae}{Vaskonen \&
  Veerm\"ae}{2021}]{Vaskonen:2020lbd}
Vaskonen V.,  Veerm\"ae H.,  2021, \mn@doi [Phys. Rev. Lett.]
  {10.1103/PhysRevLett.126.051303}, 126, 051303

\bibitem[\protect\citeauthoryear{Visinelli}{Visinelli}{2018}]{Visinelli:2017qga}
Visinelli L.,  2018, \mn@doi [Symmetry] {10.3390/sym10110546}, 10, 546

\bibitem[\protect\citeauthoryear{Visinelli \& Gondolo}{Visinelli \&
  Gondolo}{2015}]{Visinelli:2015eka}
Visinelli L.,  Gondolo P.,  2015, \mn@doi [Phys. Rev. D]
  {10.1103/PhysRevD.91.083526}, 91, 083526

\bibitem[\protect\citeauthoryear{{Wagoner}, {Fowler}  \& {Hoyle}}{{Wagoner}
  et~al.}{1967}]{1967ApJ...148....3W}
{Wagoner} R.~V.,  {Fowler} W.~A.,   {Hoyle} F.,  1967, \mn@doi [Astrophys. J.]
  {10.1086/149126}, \href
  {https://ui.adsabs.harvard.edu/abs/1967ApJ...148....3W} {148, 3}

\bibitem[\protect\citeauthoryear{Xu, Gong  \& Zhang}{Xu
  et~al.}{2020}]{Xu:2020jpv}
Xu Z.,  Gong X.,   Zhang S.-N.,  2020, \mn@doi [Phys. Rev. D]
  {10.1103/PhysRevD.101.024029}, 101, 024029

\makeatother
\end{thebibliography}

\appendix
\section{Accretion prior matter-radiation equality}
\label{sec:Sudden-Accretion}

In this Appendix, we analyse the form of the WIMP halo expected to form around a PBH, with special emphasis on the central region. This problem has been analysed before but in different contexts, so it is interesting to clarify the relationship between these previous studies. Since the initial profile in the central region is ultimately hidden by the effects of WIMP annihilations, these considerations have little impact on the constraints on the PBH and WIMP masses, so this discussion is relegated to the Appendix.
 
There are two situations, depending on whether the WIMP's kinetic energy is larger or smaller than the potential energy associated with the gravitational field of the PBH at turn-around~\citep{Adamek:2019gns}. Equation~\eqref{eq:turnaround} for $r_{\rm ta}$ neglects the kinetic energy and only applies outside the radius $r_{\Krm}$ given by Eq.~\eqref{eq:rKdef}. This leads to an $r^{-9/4}$ density profile. Within $r_{\Krm}$ the kinetic energy dominates and this tends to prevent the formation of a bound halo. However, the WIMP distribution still contains a low-velocity tail and this leads to an $r^{-3/2}$ profile in the central regions~\citep{Eroshenko:2016yve}, this being the key signature of the central black hole. Another difference is that after turn-around the WIMPs tend to move inwards in the first situation and outwards in the second, leading to an increase and decrease in the velocity dispersion, respectively. The analysis of these two situations is somewhat different, as we now describe.

The gravitational field of a PBH leads to a concentration in the distribution of surrounding WIMPs~\citep{Ullio:2001fb}. Assuming phase-space conservation, the WIMP density around the PBH is~\citep{Eroshenko:2016yve}
\begin{equation}
	\label{eq:rhoEroshenkoA}
	\rho ( r )
		=
					\frac{ 2 }{ r^{2} }\.\int \d^{3}{\bm v_{i}}\;f( {\bm v_{i}} )
					\int_{1}^{+\infty} \d r_{i}\;r_{i}^{2}\,
					\frac{ \tilde \rho_{i}( r_{i} ) }{ \tau_{\rm orb} }\!
					\(
						\frac{ \d t }{ \d r }
					\)
					,
\end{equation}
where $r$ is the current distance of the WIMP from the PBH, $r_{i}$ is its distance at turn-around, $\tilde \rho_{i}( r_{i} )$ is the WIMP density profile if one neglects kinetic energy, and ${\bm v_{i}}$ is the WIMP velocity (assuming it is bound to the PBH). The velocity distribution function $f( {\bm v_{i}} )$ is normalised so that $\smallint \d^{3}{\bm v_{i}}\,f( {\bm v_{i}} ) = 1$.
We assume this has the form
\begin{equation}
	f( v )
		=
					\frac{ 1 }{ ( 2 \pi \sigma^{2} )^{3/2} }\.
					\exp\!
					\left(
						- \frac{ v^{2} }{ 2\.\sigma^{2} }
					\right)
					,
					\label{eq:maxwell}
\end{equation}
where the velocity dispersion $\sigma$ is assumed to be isotropic. We normalise radii to $r_{\Srm} = 2\.G M$ by setting
\begin{align}
	x
		&\equiv
					r / r_{\Srm} \, ,
	\quad
	x_{i}
		\equiv
					r_{i} / r_{\Srm}
					\; .
					\label{eq:x-and-xi-definitions}
\end{align}
Although the halo is continually growing, the density at fixed $r_{i}$ is constant, so we can take the `initial' halo profile $\tilde \rho_{i}( r_{i} )$ to be the profile at $t_{\rm eq}$, as given by Eq.~\eqref{eq:WIMPprofile-ini}.

From energy conservation, the orbital period and radial speed are
\begin{subequations}
\begin{align}
	\tau_{\rm orb}
		&=
					\pi\,r_{\Srm}\,z^{3/2}
					\, ,
					\label{eq:tauorb}
					\\[1.5mm]
	\frac{ \d t }{ \d r }
		&=
					\left[
						\frac{ 1 }{ x }
						-
						\frac{ 1 }{ z }
						-
						\(
							\frac{ x_{i}\.v_{i} }{ x }
						\)^{\!2}
						\(
							1
							-
							y^{2}
						\)
					\right]^{-1/2}
					\label{eq:drdt}
					,
\end{align}
\end{subequations}
where
\begin{equation}
	v_{i}
		=
					| {\bm v_{i}} |
					\, , \quad
	z
		\equiv
					x_{i} / ( 1 - x_{i}\.v_{i}^{2} )
					\, , \quad
	y
		=
					\cos\theta
					\,
\end{equation}
and $\theta$ is the angle between $v_{i}$ and $r_{i}$, so that the angular momentum is $l = m_{\chi}\.r_{i} v_{i} \sin\theta$. We can then write integral~\eqref{eq:rhoEroshenkoA} as~\citep{Boucenna:2017ghj}
\begin{equation}
	\rho( x )
		=
					\frac{ 4 }{ x }
					\int \d v_{i}\;v_{i} f( v_{i} )\!
					\int \d x_{i}\;\frac{ x_{i}\. \tilde \rho_{i}( x_{i} ) }{ z^{3/2} }\!
					\int\mspace{-2mu}\frac{ \d y }{ \sqrt{y^{2} + y_{\mrm}^{2}\,} }
					\, ,
					\vphantom{_{_{_{_{_{_{_{_{_{_{_{_{_{_{_{_{_{}}}}}}}}}}}}}}}}}}
					\label{eq:rhoEroshenkor}
\end{equation}
where 
\begin{equation}
	y_{\mrm}^{2}
		\equiv
					\!\(
						\frac{ x }{ x_{i}\.v_{i} }
					\)^{\!2}\!
					\(
						\frac{ 1 }{ x }
						-
						\frac{ 1 }{ z }
					\)
					-
					1
		\equiv
					\zeta_{\mrm}^{2}
					-
					1
					\, 
\label{zeta}
\end{equation}
with $\zeta_{\mrm} \geq 1$. The range of $y$-integration is $1 \geq | y | \geq 0$ and performing this integral gives 
\begin{equation}
	\rho( x )
		=
					\frac{ 4 }{ x }
					\int\d v_{i} v_{i} f( v_{i} )
					\int\d x_{i}
					\frac{ x_{i}\.\tilde \rho_{i}( x_{i} ) }{ z^{3/2} }\,
					\ln\!
					\left(
						\frac{ \zeta_{\mrm} + 1 }{ \zeta_{\mrm} - 1 }
					\right)
					.
					\vphantom{_{_{_{_{_{_{_{_{_{_{_{_{_{_{_{_{_{}}}}}}}}}}}}}}}}}}
					\label{eq:rhoEroshenkor1}
\end{equation}
As discussed later, care is required if $\zeta_{\mrm} = 1$ within the range of integration since the logarithmic term then diverges. 

The region of $( x_{i}, v_{i} )$ integration is indicated in Fig.~\ref{fig:intrange}. For each value of $x_{i}$ we first integrate over $v_{i}$, where the velocity range is derived as follows. Demanding that $z$ in Eq.~\eqref{eq:tauorb} be positive yields $v_{i}^{2} < 1 / x_{i}$ and this is equivalent to the particle being gravitationally bound to the black hole since the total energy is
\begin{equation}
	E
		=
					\frac{ 1 }{ 2 }\.v_{i}^{2}
					-
					\frac{ G M }{ r_{i} }
		=
					\frac{ 1 }{ 2 }\mspace{-2mu}
					\left(
						v_{i}^{2}
						-
						\frac{ 1 }{ x_{i} }
					\right)
					.
\end{equation}
Only some fraction of the particles may satisfy this condition at turn-around, so this excludes a large fraction of the $( x_{i}, v_{i} )$ domain in Fig.~\ref{fig:intrange}. The requirement that the expression under the square root in Eq.~\eqref{eq:drdt} be positive at $y^{2} = 1$ and $y^{2} = 0$ gives
\begin{subequations}
\begin{align}
	&\frac{ 1 }{ x }
	-
	\frac{ 1 }{ x_{i} }
	+
	v_{i}^{2}
		\geq
					0
					\, ,
					\label{eq:bound1}\\[1mm]
	&\frac{ 1 }{ x }
	-
	\frac{ 1 }{ x_{i} }
	+
	\left(
		1
		-
		\frac{ x_{i}^{2} }{ x^{2} }
	\right)
	v_{i}^{2}
		\geq
					0
					\, ,
					\label{eq:bound2}
\end{align}
\end{subequations}
respectively. This implies the following bounds:
\begin{subequations}
\begin{alignat}{2}
	\label{eq:bounds}
	&\frac{ x }{ x_{i}\.( x + x_{i} ) }
		\leq
					v_{i}^{2}
		\leq
					\frac{ 1 }{ x_{i} }
					\qquad
					&\hbox{for $x_{i} \leq x$}
					\, ,
					\\[1mm]
	&0
		\leq
					v_{i}^{2}
		\leq
					\frac{ x }{ x_{i}\.( x + x_{i} )}
					\qquad
					&\hbox{for $x_{i} > x$}
					\, .
\end{alignat}
\end{subequations}
In evaluating integral \eqref{eq:rhoEroshenkor} we must therefore distinguish between regions with $x_{i} \leq x$ (lower white region) and $x_{i} > x$ (upper white region). For each of these regions we evaluate the integral numerically by slicing the portion shown in Fig.~\ref{fig:intrange} horizontally between the allowed velocity bounds.\footnote{The code used for this work is publicly available at \href{https://github.com/lucavisinelli/WIMPdistributionPBH}{github.com/lucavisinelli/WIMPdistributionPBH.}}

\begin{figure}
	\includegraphics[width = 0.95\linewidth]{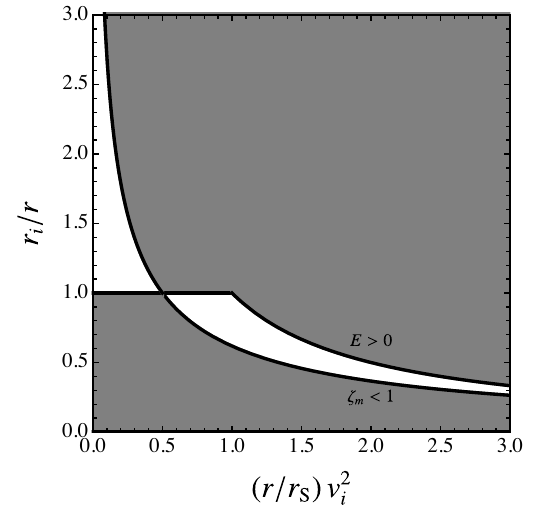} 
	\caption{Boundaries of the allowed region of integration
			for the integral in Eq.~\eqref{eq:rhoEroshenkor1}. 
			A more detailed description can be found in text of this appendix.}
	\label{fig:intrange}
\end{figure}

The lower white region in Fig.~\ref{fig:intrange} applies if the kinetic energy dominates the potential energy at turn-around, since this will cause the particle to move outwards ($r_{i} \leq r$). We then write Eq.~\eqref{eq:rhoEroshenkor1} as
\begin{align}
	\label{eq:rhoEroshenko-Kin}
	\rho( x )
		=
					\frac{ 2 }{ x }
					\int_{1}^{x}\! \d x_{i}\;
					\frac{\tilde \rho_{i}( x_{i} ) }{ x_{i}^{1/2} }\!
					\int_{\frac{ x }{ x_{i}( x + x_{i} ) }}^{\frac{ 1 }{ x_{i} }}\!
					\d w\;
		&f( w )\mspace{1mu}
					( 1 - x_{i}\.w )^{3/2}
					\notag
					\\[1mm]
		&\times
					\ln\!
					\left(
						\frac{ \zeta_{\mrm} + 1 }{ \zeta_{\mrm} - 1 }
					\right)\,,
\end{align}
where $w \equiv v_{i}^{2}$ and the bounds in the $w$ integral correspond to Eq.~\eqref{eq:bounds}. The lower limit in the $x$-integration is $1$ since $r$ must exceed $r_{\Srm}$. If we neglect the dependence of the logarithmic function in Eq.~\eqref{eq:rhoEroshenko-Kin} on $w$ (which is correct to 1st order) and use
\begin{align}
	\int_{W}\d w\;\frac{ e^{-\frac{ w }{ 2\mspace{1mu}\sigma^{2} }} }
	{ ( 2 \pi\.\sigma^{2} )^{3/2} }\.
	( 1 - w\.x_{i} )^{3/2}
		&=
					\frac{i\,\sqrt{32\,}}{(2\pi)^{3/2}}\.
					e^{-\frac{ 1 }{ 2\mspace{1mu}x_{i} \sigma^{2} }}\.
					x_{i}^{3/2}\.
					\notag
					\\[1mm]
		&\mspace{21mu}
					\times
					\sigma^{2}\,
					\Gamma\!
					\left( 
						\frac{ 5 }{ 2 },\.
						\frac{ W x_{i} - 1 }{ 2\mspace{2mu}x_{i}\.\sigma^{2} }
					\right)
					,
					\label{eq:wint}
\end{align}
then Eq.~\eqref{eq:rhoEroshenko-Kin} gives
\begin{align}
	\label{eq:rhoEroshenko-Kin1}
	\rho( x )
		&\approx
					\frac{ i\,16\.\sqrt{2\,} }{ (2\pi)^{3/2}\.x }
					\int_{1}^{x} \d x_{i}\;
					x_{i}\.\tilde \rho_{i}( x_{i} )\,
					e^{-\frac{ 1 }{ 2 x_{i} \sigma^{2} }}\.\sigma^{2}
					\nonumber
					\\[1.5mm]
		&\mspace{20mu}
					\times\!
					\left[
						\Gamma\!
						\left(
							\frac{ 5 }{ 2 },\.- \frac{ 1 }{ 2( x + x_{i} )\.\sigma^{2} }
						\right)
						-
						\Gamma\!
						\left(
							\frac{ 5 }{ 2 }, 0
						\right)
					\right]
					,
\end{align}
where the justification for dropping the logarithmic term will be given later. The Gamma functions can be expanded as
\begin{align}
	\label{eq:expand}
	\Gamma\!
					\left(
						\frac{ 5 }{ 2 },\.- \frac{ 1 }{ y }
					\right)
		&=
					\frac{ 3\sqrt{\pi\,} }{ 4 } - \frac{ 2\.i }{ 5\.y^{5/2} }
					+
					\Ocal\!
					\left(
						y^{-7/2}
					\right)				
\end{align}
for large $y$ (i.e.\ for $x \gg 1 / \sigma^{2}$) and this is just the condition that the kinetic energy much exceeds the potential energy. The constant term cancels in Eq.~\eqref{eq:rhoEroshenko-Kin1}, so we obtain
\begin{equation}
	\label{eq:rhoEroshenko-Kin2}
	\rho( x )
		\approx
					\frac{ 32\.\sqrt{2\,} }{ 5\.x^{7/2} }
					\int_{1}^{x} \d x_{i}\;x_{i}\,
					\frac{ \tilde \rho_{i}( x_{i} ) }
					{ (2\pi\sigma^{2})^{3/2} }\.
					\left(
						1
						+
						\frac{ x_{i} }{ x }
					\right)^{\!-5/2}
					e^{-\frac{ 1 }{ 2 x_{i} \sigma^{2} }}
					\, .
\end{equation}
The penultimate term introduces a numerical factor of $( 8 - 5 \sqrt{2\,} ) / 3 \approx 0.31$ in the integral and we discuss the effect of the neglected logarithmic term later.
 
We calculate integral \eqref{eq:rhoEroshenko-Kin2} in two separate domains: (i) $r < r_{\rm ta}(t_{\rm KD})$; (ii) $r_{\rm ta}(t_{\rm KD}) < r < r_{\Krm}$. There is also a third domain, (iii) $r > r_{\Krm}$, where Eq.~\eqref{eq:rhoEroshenko-Kin2} is inapplicable, and we discuss this later. We first note that $\sigma$ itself depends upon $r_{i}$ since it is determined by the temperature $T$ when the turn-around radius is $r_{i}$:
\begin{equation}
	\sigma ( r_{i} )
		\propto
					T^{1/2}
		\propto
					t^{-1/2}
		\propto
					r_{i}^{-3/4}
		\quad
		\Rightarrow
		\quad
					x_{i}\.\sigma^{2}
		\propto
					x_{i}^{-1/2}
					\,,
\end{equation}
where we have used Eq.~\eqref{eq:turnaround}. In region (i), $\tilde \rho_{i}$ and $\sigma$ are constant, so the condition $x_{i} > 1 / \sigma^{2}$ corresponds to a {\it lower} limit on $r_{i}$ and this corresponds to the radius $r_{\Crm}$ given by Eq.~\eqref{eq:C}. The exponential term gives an effective lower integral cut-off at this value. However, this cut-off can be neglected for $x > 1 / \sigma^{2}$ and this is required for the validity of Eq.~\eqref{eq:expand}. In region (ii), $\tilde \rho_{i} \propto r_{i}^{-9/4}$ and $\sigma \propto r_{i}^{-3/4}$, so the condition $x_{i} > 1 / \sigma^{2}$ corresponds to an {\it upper} limit on $r_{i}$ and this corresponds to the scale $r_{\Krm}$ given by Eq.~\eqref{eq:rKdef}. But this upper limit exceeds $x$ in domain (ii), so can be neglected. In both cases, the $r_{i}$ dependence in the ratio $\tilde \rho_{i} / ( \sigma^{2} )^{3/2}$ cancels, so we have
\begin{align}
	\label{eq:rhoEroshenko-RegI}
	\rho( x )
		\approx
					\frac{ 16\.\sqrt{2\,}\.\rho_{\rm KD} }{ 5\,x^{3/2} }
					\left(
						\frac{ m_{\chi} }{ 2 \pi\.T_{\rm KD}}
					\right)^{\!3/2}
					.
\end{align}
This explains why the low-velocity tail of the WIMP distribution generates an $r^{-3/2}$ profile at small $r$.\footnote{The $r^{- 3/2}$ behaviour seems to have been first derived by \cite{1972GReGr...3...63P}, although for stars rather than WIMPs. His derivation assumes adiabatic growth and essentially uses Eq.~\eqref{eq:rhoEroshenkoA}.} We note that the density exceeds $\rho_{\rm KD}$ for $r$ less than the value $r_{\Crm}$ given by Eq.~\eqref{eq:C} and this corresponds to $x < 1 / \sigma^{2}$, which violates the starting assumption. Thus the $r^{-3/2}$ profile is only generated in the outer constant-density region ($r > r_{\Crm}$). We return to this issue of what happens in the inner part later.

The above analysis assumes that the variation in the logarithmic term can be neglected. Since Eq.~\eqref{zeta} implies
\begin{equation}
	\zeta_{\mrm}^{2}
		=
					\left(
						\frac{ x }{ x_{i} }
					\right)^{\!2} 
					-
					\frac{ x\.( x - x_{i} ) }{ x_{i}^{3}\,w } 
					\, ,
					\label{eq:zeta2}
\end{equation}
one expects $\zeta_{\mrm} \rightarrow 1$ near the upper limit of the $x_{i}$ integral in Eq.~\eqref{eq:rhoEroshenko-Kin} ($x_{i} = x$). This would cause the logarithmic term to diverge according to
\begin{align}
	\ln\!
	\left(
		\frac{ \zeta_{\mrm} + 1 }{ \zeta_{\mrm} -1 }
	\right)\!
		\approx
					\ln\!
					\left(
						\frac{ x_{i} }{ x - x_{i} }
					\right)
					,
					\label{eq:M6}
\end{align}
but not the integral itself. If we neglect the last two terms in Eq.~\eqref{eq:rhoEroshenko-Kin2} and recall that $\tilde \rho_{i} / ( \sigma^{2} )^{3/2}$ is constant, the $x_{i}$ integral becomes 
\begin{align}
\begin{split}
	&\int_{1}^{x}\d x_{i}\;
	2\.x_{i}
	\ln\!
	\left(
		\frac{ x_{i} }{ x - x_{i} }
	\right)
	\\[1mm]
	&\qquad
		=
					x^{2}\.( 1 + \ln x )
					-
					x
					-
					( x^{2} - 1 )\.\ln( x - 1 )
					\, .
\end{split}
\end{align}
This just goes as $x^{2}$ for $x \gg 1$, which this does not alter the $x^{-3/2}$ behaviour of Eq.~\eqref{eq:rhoEroshenko-RegI}.

We now consider the case in which the kinetic energy is initially smaller than the potential energy. This applies for $r < r_{\Crm}$ and $r > r_{\rm K}$. In both cases, one expects $r < r_{i}$ and so transfers to the upper white region in Fig.~\ref{fig:intrange}.\footnote{The argument that gravity reduces the distance ($r < r_{i}$) is a statistical one and need not apply for each individual WIMP: most particles in this case spend most of the time closer to the PBH than their turn-around radius and very few will move further away.} We first consider the $r > r_{\rm K}$ situation. The lower and upper integral limits in Eq.~\eqref{eq:rhoEroshenko-Kin} now become $0$ and $x / [ x_{i} ( x + x_{i} ) ]$, respectively, while the $x_{i}$ integral has a lower (rather than upper) limit at $x$, so we have
\begin{align}
	\label{eq:rhoEroshenk}
	\rho( x )
		=&
					\frac{ 2 }{ x }
					\int_{x}^{\infty} \d x_{i}\;
					\frac{ \tilde \rho_{i}( x_{i} ) }{ x_{i}^{1/2} }\!
					\int_{0}^{\frac{ x }{ x_{i}( x + x_{i} ) }}
					\d w\;
					\frac{ 1 }{ ( 2 \pi\.\sigma^{2} )^{3/2} }\.
					e^{-w/(2 \sigma^{2})}
					\notag
					\\[1.5mm]
		&\times
					( 1 - x_{i}\.w )^{3/2}\.
					\ln\!
					\left(
						\frac{ x
							\big(
								\frac{ 1 }{ x }
								-
								\frac{ 1 }{ x_{i} }
								+
								w
							\big)^{1/2}
							+
							x_{i}\. w^{1/2}}
							{ x
								\big(
									\frac{ 1 }{ x }
									-
									\frac{ 1 }{ x_{i} }
									+ 
									w
								\big)^{1/2}
								-
								x_{i}\. w^{1/2}}
					\right)
					.
\end{align}
Because $w \rightarrow 0$ at the integral limit $x_{i} = x$, one no longer has $\zeta_{\mrm} \rightarrow 1$ in this case, so the logarithmic term must be included.

The evaluation of integral \eqref{eq:rhoEroshenk} is complicated but the analysis is simplified if one follows \cite{Adamek:2019gns} in assuming a delta-function velocity distribution, $f( v_{i} ) \propto \delta( v_{i} )$, in Eq.~\eqref{eq:rhoEroshenkoA}. There are then two ways to do the calculation. The first is to use the 3-dimensional delta function $\delta^{(3)}( \bf{v_{i}} )$ in Eq.~\eqref{eq:rhoEroshenkoA} and put $v_{i} = 0$ in the expressions for $z$ and $\d t / \d r$, this being contained within the upper white region in Fig.~\ref{fig:intrange}. The WIMP density around the PBH then becomes
\begin{equation}
	\label{eq:rhoEroshenkocc}
	\rho( r )
		=
					\frac{ 2 }{ \pi }\.
					\int_{r}^{\infty} \d r_{i}\;
					\frac{ r_{i} }{ r^{3/2} }\.
					\frac{ \tilde \rho_{i}( r_{i} ) }
					{ \sqrt{r_{i} - r\,} }
					\, ,
\end{equation}
where $\tilde \rho_{i}( r_{i} ) \equiv \rho_{\rm KD}\.( r_{\Mrm} / r_{i} )^{9/4}$ and $r_{\Mrm} \equiv ( r_{\Srm}\, t_{\rm KD}^{2})^{1/3}$ is the turn-around radius at KD.
This gives
\begin{align}
\begin{split}
	\rho( r )
		&=
					\frac{ 2\.\rho_{\rm KD} }{ \pi }
					\(
						\frac{ r_{\Mrm} }{ r }
					\)^{\!9/4}
					\int_{1}^{\infty}\d \xi \;
					\frac{ \xi_{i}^{-5/4} }
					{ \sqrt {\xi_{i} - 1\,} }
					\\[1.5mm]
		&=
					\alpha_{\Erm}\,\rho_{\rm KD}
					\(
						\frac{ r_{\Mrm} }{ r }
					\)^{\!9/4}
					,
					\label{eq:alphaE}
\end{split}
\end{align}
where $\xi \equiv r_{i} / r$ and 
\begin{equation}
	\alpha_{\Erm}
		=
					\frac{8\,\Gamma( 3/4) }
					{ \sqrt{\pi\,}\,\Gamma( 1/4 ) }
		\approx
					1.53
					\, . 
\end{equation}
Therefore, the only effect of the kinetic energy with this approximation is to increase the density everywhere by a factor $\alpha_{\Erm}$, as pointed out by~\cite{Adamek:2019gns}. 

The second approach relates more directly to the previous analysis and represents the 3-dimensional delta function as
\begin{equation}
	\delta( v_{i} )
		=
					\lim_{\sigma \rightarrow 0}
					\left[
						\frac{ 4\pi\.v_{i}^{2} }{ ( 2 \pi\.\sigma^{2} )^{3/2} }\.
						e^{- v_{i}^{2} / ( 2 \sigma^{2} )}
					\right]
					.
\end{equation}
In this limit, Eq.~\eqref{eq:rhoEroshenk} becomes
\begin{align}
	\label{eq:rhoEroshenk1}
	\rho( x )
		&=
					\frac{ 2 }{ x }
					\int_{x}^{\infty} \d x_{i}\;
					\frac{ \tilde \rho_{i}( x_{i} ) }{ x_{i}^{1/2} }\!
					\int_{0}^{\frac{ x }{ x_{i}( x + x_{i}) }}\d v_{i}\;
					\frac{ 1 }{ 2\pi\.v_{i} }\.\delta( v_{i} )
					\notag
					\\[1.5mm]
		&\quad\.
		\times
					( 1 - x_{i}\.v_{i}^{2} )^{3/2}
					\ln\!
					\left(
						\frac{x
							\big(
								\frac{ 1 }{ x }
								-
								\frac{ 1 }{ x_{i} }
								+
								v_{i}^{2}
							\big)^{\!1/2}
							+
							x_{i}\.v_{i} }
							{ x
								\big(
									\frac{ 1 }{ x }
									-
									\frac{ 1 }{ x_{i} }
									+
									v_{i}^{2}
								\big)^{\!1/2}
								-
								x_{i}\.v_{i} }
					\right)
					.
\end{align}
The integration over $\d v_{i}$ imposes $v_{i} = 0$ in the square-root terms, so the logarithmic function can be approximated as
\begin{equation}
\label{eq:approxzeta}
	\ln\!
	\left(
		\frac{ x
			\big(
				\frac{ 1 }{ x }
				-
				\frac{ 1 }{ x_{i} }
			\big)^{\!1/2}
			+
			x_{i}\.v_{i} }
			{ x
				\big(
					\frac{ 1 }{ x }
					-
					\frac{ 1 }{ x_{i} }
				\big)^{\!1/2}
			-
			x_{i}\.v_{i} }
	\right)\!
		\approx
					\frac{ 2\.v_{i}\.x_{i}^{3/2} }
					{ x^{1/2}\.\sqrt{x_{i} - x\,} }
					\, .
\end{equation}
Thus Eq.~\eqref{eq:rhoEroshenk1} just reduces to Eq.~\eqref{eq:rhoEroshenkocc}, as expected. This shows the important role of the logarithmic term in this case, whereas it has little effect when the kinetic energy dominates.

We can analyse the situation more precisely by dropping the assumption $\sigma \rightarrow 0$ but retaining the approximation \eqref{eq:approxzeta}. Then Eq.~\eqref{eq:rhoEroshenk} takes the form:
\begin{align}
\begin{split}
	\label{eq:rhoEroshenk3}
	\rho( x )
		&=
					\frac{ 2 }{ \pi }
					\int_{x}^{\infty} \d x_{i}\;
					\frac{ x_{i} }{ x^{3/2} }
					\frac{ \tilde \rho_{i}( x_{i} ) }{ \sqrt{x_{i} - x\,} }
					\\[1mm]
		&\mspace{90mu}
		\times
					\left[
						1 - \frac{ 2 }{ \sqrt{\pi\,} }\.
						\Gamma\!
						\left(
							\frac{ 3 }{ 2 }, 
							\frac{ x }{ 2\.\sigma^{2}\.x_{i}\.( x + x_{i} )}
						\right) 
					\right]
					,
\end{split}
\end{align}
where $\tilde \rho_{i}( x_{i} ) \propto x_{i}^{-9/4}$. The Gamma term is associated with the factor $v_{i}$ in Eq.~\eqref{eq:approxzeta} and for $\sigma^2x_i \ll 1$ it can be approximated as 
\begin{equation}
	\Gamma\!
	\left(
		\frac{ 3 }{ 2 }, \frac{ x }{ 2\.\sigma^{2}\.x_{i}\.( x + x_{i} ) }
	\right)
		\approx
					\left[
						\frac{ x }{2 \sigma^{2}\.x_{i}( x + x_{i} ) }
					\right]^{1/2}\.
					e^{- \frac{ x }{ 2 \sigma^{2} x_{i} ( x + x_{i} ) }}
					\, .
\end{equation}
We have neglected the $( 1 - x_{i}\.w )^{3/2}$ factor in Eq.~\eqref{eq:rhoEroshenk} since this is a smaller correction. In the limit $\sigma \to 0$, the Gamma function goes to zero and Eq.~\eqref{eq:rhoEroshenkocc} is again obtained.

Finally, we consider the solution of Eq.~\eqref{eq:rhoEroshenkoA} for $r < r_{\Crm}$, where $x \, \sigma^2$ is also small. In this case Eq.~\eqref{eq:rhoEroshenk3} leads to
\begin{align}
\begin{split}
	\label{eq:rho-Theta}
	\rho( \Theta )
		&=
					\int_{\Theta}^{\infty} \d y\;
					\kappa( \Theta, y )
					\, ,\vphantom{\Bigg|_{_{1}}}
\end{split}
\end{align}
where
\begin{align}
	\kappa( \Theta, y )
		&\equiv
					\frac{ 2 }{ \pi }\.
					\frac{ y }{ \Theta^{3/2} }
					\frac{ 1 }{ \sqrt{y - \Theta\,} }
					\left[
						1 - \frac{ 2 }{ \sqrt{\pi\,} }\.
						\Gamma\!
						\left(
							\frac{ 3 }{ 2 }, 
							\frac{ \Theta }{ 2\.\sigma^{2}\.y\.( \Theta + y )}
						\right) 
					\right]_{_{_{_{_{_{}}}}}}
					\!\!
					\label{eq:kappa}
\end{align}
and we have introduced the parameter $\Theta \equiv x\.\sigma^{2}$. We assume $\tilde \rho_{i}$ is constant within $r_{\Crm}$ and, since $\sigma^{2}$ is of order $10^{-4}$ for the expected WIMP parameters, $\Theta$ is necessarily small. We can then expand Eq.~\eqref{eq:rho-Theta} as
\begin{align}
	\rho( \Theta )
		&=
					\rho( 0 )
					+
					\Theta\,\frac{ \partial \rho( \Theta ) }
					{ \partial \Theta }\bigg|_{\Theta\mspace{1mu}=\mspace{1mu}0}
					+
					\Ocal\big( \Theta^{2} \big)
					\nonumber
					\\[2mm]
		&=
					\rho( 0 )
					+
					\Theta\!
					\left[
						-\.
						\kappa( 0, 0 )
						+\!
						\int_{\Theta}^{\infty}\!\d y\;
						\frac{ \partial \kappa( \Theta, y ) }
						{ \partial \Theta }\,\Big|_{\Theta\mspace{1mu}=\mspace{1mu}0}
					\right]\!
					+
					\Ocal\big( \Theta^{2} \big)
					\. . 
\end{align}				
In the context of Eq.~\eqref{eq:rhoEroshenk3}, $y$ corresponds to $x_{i}\.\sigma^{2}$, so 
\begin{equation}
		\rho (x) \simeq
					\int_{0}^{\infty}\d y\;
					\sqrt{\frac{ 2 }{ \pi }\,}\.
					\frac{ e^{- \frac{ \Theta }{ 2 y^{2} }} }{ y^{5/2} }
		=
					\left(
						\frac{ 2 }{ \pi^{2} }
					\right)^{\! 1/4} \,
					\Gamma\big( \tfrac{ 3 }{ 4 } \big)\;
					\sigma^{- 3/4} x^{-3/4}
					\; .
					\label{eq:rho-Theta-expansion}
\end{equation}
This yields an $r^{-3/4}$ power-law within $r_{\Crm}$ and this is confirmed by the numerical results in Fig.~\ref{fig:rhochi}. We note that the $\Gamma$ term in Eq.~\eqref{eq:kappa} only leads to a small correction for $r > r_{\rm K}$ but is essential for $ r < r_{\Crm}$.

\bsp
\label{lastpage}
\end{document}